\documentclass[12pt,preprint]{aastex}
\usepackage{amsmath,bm}
\usepackage{multirow}
\usepackage{enumerate}
\usepackage{graphicx}
\usepackage{color}
\usepackage[T1]{fontenc}
\usepackage{url}
\usepackage{hyperref}
\newcount\doepsf
\newcommand{\be}{\begin{equation}}
\newcommand{\ee}{\end{equation}}
\newcommand{\bea}{\begin{eqnarray}}
\newcommand{\eea}{\end{eqnarray}}
\newcommand{\bean}{\begin{eqnarray*}}
\newcommand{\eean}{\end{eqnarray*}}

 \definecolor{DarkGreen}{rgb}{0.0,0.45,0.0}     
 \definecolor{DarkMagenta}{rgb}{0.45,0.0,0.45}  

\begin{document}

\title{Blob formation and ejection in coronal jets due to the plasmoid and Kelvin-Helmholtz instabilities}

\author{Lei Ni$^{1,2}$,
         Qing-Min Zhang$^{3}$,
            Nicholas A. \ Murphy$^{4}$ and
            Jun Lin $^{1}$ 
            }

\affil{$^1$Yunnan Observatories, Chinese Academy of Sciences, 396 Yangfangwang, Guandu District, Kunming, 650216, P. R. China}
\affil{$^2$Center for Astronomical Mega-Science, Chinese Academy of Sciences, 20A Datun Road, Chaoyang District, Beijing, 100012, P. R. China}
\affil{$^3$Key Laboratory for Dark Matter and Space Science, Purple Mountain Observatory, Chinese Academy of Sciences, Nanjing 210008,  China} 
\affil{$^4$Harvard-Smithsonian Center for Astrophysics, Cambridge, Massachusetts 02138, USA}

\shorttitle{The formation of blobs in coronal jets}  
\shortauthors{Ni et al.}

\email{leini@ynao.ac.cn}

\slugcomment{CoronaJet; v.\ \today}

\begin{abstract}
\noindent  We perform two-dimensional resistive magnetohydrodynamic simulations of coronal jets driven by flux emergence along the lower boundary.  The reconnection layers are susceptible the formation of blobs that are ejected in the jet. Our simulation with low plasma $\beta$ (Case I) shows that magnetic islands form easily and propagate upwards in the jet.  These islands are multithermal and thus are predicted to show up in hot channels (335 \AA\,  and 211 \AA) and the cool channel (304 \AA) in observations by the Atmospheric Imaging Assembly (AIA) on the \emph{Solar Dynamics Observatory}. The islands have maximum temperatures of 8 MK, lifetimes of 120 s, diameters of 6 Mm, and velocities of 200 km s$^{-1}$.  These parameters are similar to the properties of blobs observed in EUV jets by AIA\@. The Kelvin-Helmholtz instability develops in our simulation with moderately high plasma $\beta$ (Case II), and leads to the formation of bright vortex-like blobs above the multiple high magneto-sonic Mach number regions that appear along the jet. These vortex-like blobs can also be identified in the AIA channels. However, they eventually move downward and disappear after the high magneto-sonic Mach number regions disappear. In the lower plasma $\beta$ case, the lifetime for the jet is shorter, the jet and magnetic islands are formed with higher velocities and temperatures, the current sheet fragments are more chaotic, and more magnetic islands are generated. Our results show that the plasmoid instability and Kelvin-Helmholtz instability along the jet are both possible causes of the formation of blobs observed at extreme ultraviolet (EUV) wavelengths.
\end{abstract}   


\section{Introduction}
\label{s:introduction}
Coronal jets are frequently observed in X-ray and extreme ultraviolet (EUV) wavelengths by  \emph{Yohkoh}, the \emph{Solar Terrestrial Relations Observatory} (\emph{STEREO}), \emph{Hinode}, the \emph{Solar Dynamics Observatory} (\emph{SDO}) and the \emph{Interface Region Imaging Spectrograph} (\emph{IRIS}). Numerous papers have described the characteristics of the observed jets \citep[e.g.,][]{1992PASJ...44L.173S,1999ApJ...513L..75C,2007PASJ...59S.771S,2008A&A...491..279C,2009SoPh..259...87N,2011ApJ...735L..43S,2012ApJ...745..164S, 2013A&A...559A...1S,2014Sci...346A.315T,2015Sci...350.1238C,2016A&A...589A..79M}. Statistical studies \citep[e.g.,][]{1996PASJ...48..123S} show that jets have characteristic lengths of $10-400$~Mm and widths of about $5-100$~Mm. The apparent velocities are about $10-1000$~km\,s$^{-1}$, with an average velocity of about $200$~km\,s$^{-1}$. The lifetimes of jets range from a few minutes to several hours. The temperature of polar jets ranges from $0.1$~MK to $10$~MK. 

Generally, jets are believed to be triggered by magnetic reconnection \citep{1995Natur.375...42Y}. Different magnetic configurations and altitudes where magnetic reconnection takes place result in different types of jets with different characteristics. As pointed out by \cite{2007Sci...318.1591S}, magnetic reconnection in the chromosphere triggers chromospheric anemone jets. Reconnection taking place in the corona results in the larger sizes, faster speeds, longer lifetimes and higher temperature in jets. \cite{2014A&A...567A..11Z} reported the first discovery of recurring blobs in homologous EUV coronal jets observed by the Atmospheric Imaging Assembly \citep[AIA;][]{2012SoPh..275...17L} on \emph{SDO}, which has  unprecedented temporal and spatial resolution. \citet{2016SoPh..291..859Z} studied similar incoherent EUV coronal jets composed of bright and compact blobs with characteristic sizes of $2$--$10$ Mm and apparent velocities of $120$--$450$~km\,s$^{-1}$. The blobs are identifiable in multiple wavelengths, and differential emission measure (DEM) analyses show multithermal dynamics with characteristic temperatures of $1.8$--$3.1$~MK\@. \citet{2017ApJ...834...79Z} recently recognized multiple bright blobs in the legs of an eruptive jet using high resolution data from \emph{IRIS} \citep[]{2014SoPh..289.2733D}. Some blobs moved upward while others moved downward to the solar surface. The high order tearing instability known as the plasmoid instability operating during turbulent magnetic reconnection is one possible mechanism for the generation of these bright blobs. Plasmoids generated during magnetic reconnection are also called magnetic islands in 2D, with one O-type magnetic null point located inside each island. For this mechanism, the observed blobs would correspond to the magnetic islands or magnetic flux ropes (3D) generated during this instability. In this paper,  "magnetic island" refers to a magnetic structure with closed field lines surrounding an magnetic O-point, and "blob" refers to a round structure with high density and high temperature that is bright in AIA (either observationally or numerically).  

Two-dimensional (2D) and three-dimensional (3D) numerical experiments have been previously performed to investigate the physical mechanisms that accelerate jets. A leading theoretical idea is that jets are launched when magnetic flux emerges from the photosphere and reconnects with pre-existing overlying flux.  \citet{1995Natur.375...42Y, 1996PASJ...48..353Y} carried out the first 2.5D magnetohydrodynamic (MHD) numerical simulations to investigate jet ejection from an inverted-Y configuration. They observed the formation of a current sheet between the emerging magnetic loops and the background magnetic field. Strong collisions between the reconnection outflow and background magnetic field lead to a fast mode shock which diverts part of the outgoing plasma upward along open fields, thus creating a jet. \cite{2008ApJ...683L..83N} extended the simulations by using more realistic plasma density and temperature. These simulations produced both hot ($\sim5$~MK) and cool ($\sim0.01$~MK) jets, which are comparable to jets observed by \emph{Hinode}.  \citet{2012ApJ...751..152J} and \cite{2013ApJ...777...16Y} presented higher resolution simulations that showed the formation of multiple magnetic islands in the current sheet region. \citet{2013ApJ...777...16Y} include heat conduction and radiative cooling in chromospheric jet simulations, and trigger magnetic reconnection using moving magnetic features rather than flux emergence. However, in all of the previous numerical simulations, the upward moving magnetic islands eventually merge into the background magnetic fields and plasma. None of the simulated magnetic islands have been observed to exist and move out along the jet as shown in the observational results by \citet{2014A&A...567A..11Z} and \citet{2016SoPh..291..859Z}.

\citet{2008ApJ...673L.211M} reported the first 3D numerical simulation of a jet driven by magnetic flux emergence from the photosphere. The magnetic skeleton and topology of their results are very similar to linear force-free extrapolations of magnetograms from the Michelson Doppler Imager (MDI) on the \emph{Solar and Heliospheric Observatory} (\emph{SOHO}), but vertical cuts of the magnetic field, current density, and temperature are all similar to prior 2D results. They later extend this work by using a larger computational domain and running the simulation for a significantly longer time \citep{2013ApJ...771...20M}. After the decay of the hot and fast inverted-Y shape jet, a violent phase with a total of five eruptions occurred. Their simulations may provide a model for blowout jets. The 3D numerical simulation by \cite{2013ApJ...769L..21A} showed the transition from ``standard'' to ``blowout''. \citet{2009ApJ...691...61P, 2015A&A...573A.130P} proposed a 3D null-point configuration to study jet activity. The energy is stored by a twisting motion until kink-like instability breaks the symmetry and leads to an explosive release of energy via reconnection. Their model may explain recurrent 3D twist jets. Previous numerical simulations and theory \citep{2000mrp..book.....B, 2009PhPl...16k2102B, 2010PhPl...17e2109N, 2013PhPl...20f1206N, 2016JPlPh..82f5901C} establish that the Lundquist number must be high enough to trigger the plasmoid instability. \cite{2006ApJ...645L.161A} and  \cite{2013ApJ...771...20M} already found that 3D magnetic magnetic flux rope were formed in the main current sheet in a flux emergence context. By using higher resolution to achieve a higher numerical Lundquist number ($\simeq2\times10^4$), \citet{2016ApJ...827....4W} observed multiple high density twist flux-rope structures in the current layer which are corresponding to the plasma blobs. They estimated some observable properties of these blobs by adopting typical values for the length scale ($L_s=10^6$\,m), field strength ($B_s=0.001$\,T), and plasma density in solar jets ($\rho_s=10^{-11}$\,kg\,m$^{-3}$). The fastest blob that existed in the current layer in their simulations had velocities around $300$~km\,s$^{-1}$ and lifetimes around $12.5\sim25$\,s, which are compatible with the characteristics of the observed bright blobs \citep{2016SoPh..291..859Z}.  However, they pointed out that as the twist in the ropes spread out along the length of their field lines, so did the region of enhanced density. The blobs were then assimilated into the higher-density regions of the current-sheet outflow. Therefore, similar to the previous 2D simulations \citep[]{2013ApJ...777...16Y}, they did not show the clear density enhanced blobs in the jet region. In addition, radiative cooling and heat conduction are not included to study the thermal structures in jets in these 3D simulations.

We perform high resolution 2D MHD simulations of coronal jets to achieve a Lundquist number that is high enough for the plasmoid instability to develop. Our simulations are the first to show upward ejected blobs with multithermal dynamics that move outward along the jet.  The jets are host to fast mode shocks and many chaotic current sheet fragments.  Section \ref{s:model} describes our numerical model and simulation setup.  We present our numerical results in Section \ref{s:results}.  A summary and discussion are given in Section \ref{s:discussion}. 

\section{Numerical model and initial conditions}
\label{s:model}
The single-fluid MHD equations  with gravity, radiative cooling, heating, and heat conduction that we use in our simulations are:

\begin{eqnarray}
 \partial_t \rho &=& -\nabla \cdot \left(\rho \mathbf{v}\right),                                                                \\ 
 \partial_t \mathbf{B} &=& \nabla \times \left(\mathbf{v} \times \mathbf{B}-\eta\nabla \times \mathbf{B}\right), \label{e:induction}\\
  \partial_t (\rho \mathbf{v}) &=& -\nabla \cdot \left[\rho \mathbf{v}\mathbf{v}
                              +\left(p+\frac {1}{2\mu_0} \vert \mathbf{B} \vert^2\right)\mbox{\bfseries\sffamily I} \right]  \nonumber \\
                              & &+\nabla \cdot \left[\frac{1}{\mu_0} \mathbf{B} \mathbf{B} \right] + \rho \mathbf{g}, \\
 \partial_t e &=& - \nabla \cdot \left[ \left(e+p+\frac {1}{2\mu_0 }\vert \mathbf{B} \vert^2\right)\mathbf{v} \right] \nonumber\\
       & &+\nabla \cdot \left[\frac {1}{\mu_0} \left(\mathbf{v} \cdot \mathbf{B}\right)\mathbf{B}\right] + \nabla \cdot \left[ \frac{\eta}{\mu_0} \textbf {B} \times \left(\nabla \times \mathbf{B}\right) \right] \nonumber\\
       & & -\nabla \cdot \mathbf{F}_\mathrm{C}+\rho \mathbf{g} \cdot \mathbf{v}+\mathcal{L}_\mathrm{rad}+\mathcal{H},   \\
   e &=& \frac{p}{\Gamma_0-1}+\frac{1}{2}\rho \vert  \mathbf{v} \vert^2+\frac{1}{2\mu_0}\vert \mathbf{B} \vert^2,          \\
   p &=& \frac{2\rho}{m_\mathrm{i}} k_\mathrm{B}T .
\end{eqnarray}
Here $\rho$ is the plasma mass density, $\mathbf{v}$ is the center of mass velocity, $e$ is the total energy density, $\mathbf{B}$ is the magnetic field, $\eta$ is the magnetic diffusivity, $p$ is plasma thermal pressure, $\mathbf{g}=-273.9~\mbox{m\,s}^{-2}~\mathbf{e}_y$ is the gravitational acceleration of the Sun. $\mathcal{L}_\mathrm{rad}$ is the radiative cooling function, $\mathcal{H}$ is the background heating function and $\mathbf{F}_\mathrm{C}$ is the heat conduction. The international system of units (SI) are applied for all the variables in this work.

According to the paper by \citet{1980SoPh...68..351N} and \citet{2015ApJ...812...92N}, we assume the following analytical expressions for optically thin radiative cooling $\mathcal{L}_\mathrm{rad}$ and heating $\mathcal{H}$:

\begin{eqnarray}
    \mathcal{L}_\mathrm{rad} &=&
        \begin{cases}
            2.23872 \times 10^{-27} a(\rho_0)\left(\frac{\rho}{m_\mathrm{i}}\right)^2T^{-1.385} & 
            \textrm{$2.5 \times 10^{5}$~K $< T <$ $10^{6}$~K}\\
            4.64515 \times 10^{-32} a(\rho_0) \left(\frac{\rho}{m_\mathrm{i}}\right)^2T^{-0.604} & 
            \textrm{ $10^{6}$~K $\leq T \leq$ $2\times 10^{7}$~K}
        \end{cases}, \\
       \mathcal{H} &=& 2.23872 \times 10^{-27} a(\rho_0)\frac{\rho \rho_0} {{m_\mathrm{i}}^2}T_0^{-1.385},
\end{eqnarray}
where $T_0$ is the initial temperature and $\rho_0$ is the initial density in the whole simulation domain at $t=0$. The unit for $ \mathcal{L}_\mathrm{rad}$ and $\mathcal{H}$ is J\,m$^{-3}$\,s$^{-1}$. In the paper by \citet{1980SoPh...68..351N}, the parameter $a(\rho_0)=1$. The magnitude of radiative cooling is similar to the models calculated by \cite{2009A&A...498..915D} from the Chianti atomic database and the linear approximation by \cite{2008ApJ...682.1351K}. In section 2.3.4 in the book by \cite{2014masu.book.....P}, he has pointed out that the form of the coronal heating term $\mathcal{H}$ is often assumed to be either uniform or proportional to density. In this work, we assume that the heating term is proportional to density. As the temperature increases, the radiative cooling term $\mathcal{L}_\mathrm{rad}$ will decrease and the heating term $\mathcal{H}$ might be larger than the cooling term especially inside the magnetic islands with high plasma density. We assume $a(\rho_0)=0.2$ in our simulations. According to the numerical results presented in the following section, the radiative cooling and heating terms inside the magnetic islands have been calculated to compare with $ \partial_t e$ at several different locations and times. From the calculations, we find that the heating term is indeed usually higher than the cooling term. However, both of the terms are around $10$ to $100$ times smaller than the $\partial_te$ term. Therefore, the radiative cooling and heating terms only have very small effects on our numerical results. In the future, we will try to improve models for the radiative cooling and heating.      

The anisotropic heat conduction term $\mathbf{F}_\mathrm{C}$ is given by
 \begin{eqnarray}
  \mathbf{F}_\mathrm{C}=-\kappa_{\parallel}\left(\nabla T\cdot \hat{\mathbf{B}}\right)\hat{\mathbf{B}}-\kappa_{\perp}\left[\nabla T-\left(\nabla T\cdot \hat{\mathbf{B}}\right)\hat{\mathbf{B}}\right]
\end{eqnarray}
where $\hat{\mathbf{B}}=\mathbf{B}/ \vert \mathbf{B} \vert $ is a unit vector in the direction of the magnetic field. The parallel and perpendicular conductivity coefficients, 
$\kappa_{\parallel}$ and $\kappa_{\perp}$, are given by:
\begin{eqnarray}
  \kappa_{\parallel}=\frac{1.84\times10^{-10}}{\ln \Lambda} {T}^{5/2},\\
  \kappa_{\perp} = 8.04\times10^{-33} \left(\frac{\ln \Lambda}{m_\mathrm{i}}\right)^2 \frac{\rho^2}{T^3 {\arrowvert \mathbf{B} \arrowvert}^2 } \kappa_{\parallel},
 \end{eqnarray}
 where $\ln \Lambda=30$. The unit for $\kappa_{\parallel}$ and $\kappa_{\perp}$ is J\,K$^{-1}$\,m$^{-1}$\,s$^{-1}$. The validity of the perpendicular coefficient is restricted to the strong magnetic field case. In the weak field case, the theory breaks down and heat conduction approaches isotropy. The implementation accounts for this by modifying $\kappa_{\perp}$ such that $\kappa_{\perp}=\min \left({\kappa_{\perp}, \kappa_{\parallel} }\right)$ which avoids singular behavior when $\mathbf{B}\longrightarrow0$. According to the numerical results presented in the following section, we have also calculated the heat conduction term $\nabla \cdot \mathbf{F}_\mathrm{C}$ to compare with $\partial_te$ at several different locations and times. We find that the heat conduction term in the direction parallel to the magnetic field can be the same magnitude as the $\partial_te$ term in the locations with high temperature ($>3\times10^6$~K) and strong temperature gradients. However, the perpendicular heat conduction is around nine orders of magnitude smaller than $\partial_te$. Therefore, the heat conduction term is important in the direction which is parallel to the magnetic field, e.g. along the jet. However, it has little effect inside the magnetic islands and the high temperature plasmas are confined inside the islands. We have also discussed this issue in the last section in our previous paper \citep{2015ApJ...812...92N}.    
 
 The initial uniform background magnetic field is set as $B_{x0}=-0.6b_0$ and $B_{y0}=-0.8b_0$. The initial uniform temperature is $T_0=8\times10^5$~K and the initial plasma velocity is zero. Since gravity is included, the initial stratified density can be calculated from equation (3) and equation (6) as below:
 \begin{equation}
   \rho_0=\rho_{00}\mathrm{exp}\left(-\frac{\mathrm{m_i} \mathrm{g}}{2\mathrm{k_B}\mathrm{T_0}}\mathrm{y}\right)
 \end{equation}
 where $\rho_{00}= 0.5\times1.66057\times10^{-10}$~kg\,m$^{-3}$, $m_\mathrm{i}=1.66057\times10^{-27}$~kg is the mass of hydrogen ion and $k_\mathrm{B}=1.3806\times10^{-23}$~J\,K$^{-1}$ is the Boltzmann constant. In this work, we have simulated three cases. Case~I has $b_0=0.003$~T, Case~II has $b_0=0.0015$~T, and Case~III has $b_0=0.00075$~T. The 2D simulation box as shown in Figure~1(a) is in the domain $0<x<200L_0$ and $0<y<100L_0$,  with $L_0=10^6$~m. The initial plasma $\beta$ varies with height from bottom to the top boundary in the range $0.3084> \beta_0 >0.0393$ in Case~I, $1.5216>\beta_0>0.1574$ in Case~II, and $4.9349>\beta_0>0.6296$ in Case~III\@. The initial number density ($n=\rho/m_i$) and magnetic field are shown in Figure~1. We use temperature dependent magnetic diffusivity $\eta=10^8\left(T_0/T\right)^{3/2}+10^9\left[1.-\mathrm{tanh}(\frac{y-2L0}{0.2L0})\right]$~m$^2$\,s$^{-1}$ in all the three cases. The magnetic diffusion is around 10 times higher at the bottom for $y<2L_0$ than in the higher region.  
 
 At each boundary, two extra layers with ghost grid cells are applied in the code to set boundary conditions. The last paragraph in this section will give more details about the relationship between the physical boundary in the figures and the two layers with ghost grid cells. Outflow boundary conditions are applied at the left ($x=0$) and right ($x=200L_0$). Plasma is allowed to flow out of the domain but not to flow in. {The gradient of the plasma density and thermal energy (plasma pressure) vanish. The gradient of the magnetic field component parallel to the boundary vanishes.}  A divergence-free extrapolation of the magnetic field is used for the component that is perpendicular to the boundary. The plasma is also allowed to flow out of the domain but not to flow in at the top boundary ($y=100L_0$). Since gravity is included in y direction, the thermal energy (plasma pressure) and plasma density decrease with height as $e_{th}(x,y,z)=e_{thU} \mathrm{exp}\left[\frac{\mathrm{g}\rho_U\mathrm{(y-100L_0)}}{p_U}\right]$ and $\rho(x,y,z)=\rho_U\mathrm{exp}\left[\frac{\mathrm{g}\rho_U\mathrm{(y-100L_0)}}{p_U}\right]$ in the two layers with ghost grid cells at the top boundary, where $e_{thU}$, $\rho_{U}$ and $p_U$ are the thermal energy, plasma density and pressure at the highest y position inside the simulation domain respectively, $\mathrm{g}=-273.9~\mbox{m\,s}^{-2}$. The boundary conditions for the magnetic field at the top are the same as along the left and right boundaries.  For the bottom boundary, the two layers with ghost grid cells are located below the physical bottom boundary $y=0$ as shown in Figure~1(a). The plasma density is fixed at the initial value in the two layers. The gradients of plasma velocities vanish. The magnetic fields in the bottom layer with ghost grid cells are set as below:
 
 \begin{eqnarray}
   (t \leq t_1)
    \begin{cases}
        b_{xb}=-0.6b_0+\frac{100L_0(y-y_0)b_1t}{[(x-x_0)^2+(y-y_0)^2]t_1}[\mathrm{tanh}(\frac{x-70L_0}{\lambda})-\mathrm{tanh}(\frac{x-130L_0}{\lambda})]  \\
        b_{yb}=-0.8b_0-\frac{100L_0(x-x_0)b_1t}{[(x-x_0)^2+(y-y_0)^2]t_1}[\mathrm{tanh}(\frac{x-70L_0}{\lambda})-\mathrm{tanh}(\frac{x-130L_0}{\lambda})]  \\ 
    \end{cases}
\end{eqnarray}

\begin{eqnarray}
   (t \geq t_1)
    \begin{cases}
        b_{xb}=-0.6b_0+\frac{100L_0(y-y_0)b_1}{[(x-x_0)^2+(y-y_0)^2]}[\mathrm{tanh}(\frac{x-70L_0}{\lambda})-\mathrm{tanh}(\frac{x-130L_0}{\lambda})]  \\
        b_{yb}=-0.8b_0-\frac{100L_0(x-x_0)b_1}{[(x-x_0)^2+(y-y_0)^2]}[\mathrm{tanh}(\frac{x-70L_0}{\lambda})-\mathrm{tanh}(\frac{x-130L_0}{\lambda})]   \\ 
    \end{cases}
\end{eqnarray}
where $x_0=100L_0$, $y_0=-12L_0$ and $\lambda=0.5L_0$. The distributions of the magnetic fields as shown in equation (13) and (14) are only for the ghost cells below the bottom boundary of the physical domain. The magnetic field distribution does not fulfill the divergence free condition $\nabla \cdot \mathbf{B} = 0$ inside the two ghost layers around $x=70L_0$ and $x=130L_0$. However, the divergence free condition for the magnetic field can be satisfied to high precision inside the simulation box as shown in Figure~1(a) and discussed later in this section. The high magnetic diffusion smooths the non-physical features inside the ghost layers, which then smooths the related values at the bottom boundary. 

 In order to match the initial background magnetic fields and Alfv\'en speed, we set $b_1=6\times10^{-4}$~T and $t_1=250$~s in Case~I, $b_1=3\times10^{-4}$~T and $t_1=500$~s in Case~II, and $b_1=1.5\times10^{-4}$~T and $t_1=1000$~s in Case~III. As shown in the previous work  \citep[e.g.,][]{1984SoPh...94..315F, 2000ApJ...545..524C, 2010A&A...510A.111D}, the magnetic flux emergence can be set up by changing conditions at the bottom boundary. In our cases, $B_x$ and $B_y$ are independent variables, and we directly change the value of the magnetic field in the ghost layer relating to the physical bottom boundary. This kind of flux emergence boundary condition in our cases is the same as in the paper by \cite{2012ApJ...751..152J}. The flux emergence boundary can also be set up as in the work by \citet{1996PASJ...48..353Y,2001ApJ...549..608M,2001ApJ...554L.111F}, where initial perturbations or non-equilibrium initial conditions make the system evolve. The magnetic field in the system changes with time, with some magnetic field emerging to the upper region. In 3D simulations, the authors can drive the boundary with an electric field resembling a rising flux tube \citep[e.g.,][]{2008ApJ...679..871M}. Though \cite{2013ApJ...777...16Y} used moving magnetic features rather than flux emergence to simulate jet formation, the main features of the current sheet and jet are similar in the papers by \cite{2013ApJ...777...16Y} and \cite{2012ApJ...751..152J}. Therefore, the main calculations shown in this work should not be affected by non-physical features inside the ghost layers, which are smoothed by the high magnetic diffusion. The focus of this paper is on the upward outflow region of the main current sheet and jet after flux emergence stops, rather than the flux emergence process itself.       

The computations are performed by using the MHD code NIRVANA \citep[version 3.6 and 3.8;][]{2008CoPhC.179..227Z, 2011JCoPh.230.1035Z}. The hyperbolic part of the MHD equations is solved with finite volume (FV) methods within a method of line integration framework. The resulting system of ODEs is discretized with a time explicit third order accurate Runge-Kutta method. The Second order version of the Central Upwind scheme is applied to the Euler equations with Lorentz force term combined with a constrained transport (CT) scheme for the induction equation. The electric field is computed from a genuinely 2D central upwind procedure (CCT) based on the evolution projection method.  The dissipation terms are spatially discretized within the FV framework and make use of second order finite difference approximations of the dissipative fluxes. The magnetic diffusion solver keeps the divergence free condition for the magnetic field \textbf{B}. The divergence free condition of the magnetic field is a built-in property of the scheme by virtue of a constrained transport ansatz for the induction function. The relative divergence of the magnetic field which has been tested is normally smaller than $10^{-6}$. Detailed descriptions of this scheme are presented in the paper by \cite{2011JCoPh.230.1035Z}. In that paper, numerical experiments illustrate the overall robustness and performance of the scheme for some tests. 

Adaptive mesh refinement is used in this work. The details about the derivatives-based mesh refinement criterion have been described by \citet{2015ApJ...812...92N} and in the user guide file of this code. In this work, we still choose the magnetic field to set the criterion. The threshold parameter $\varepsilon_U$ for the magnetic field is set to $0.39$. The reference value  $U_{ref}$ is $3\times10^{-4}$~T for Case~I, $1.5\times10^{-4}$~T for Case~II and $0.75\times10^{-4}$ for Case~III. We start our simulations from a base-level grid of $320\times160$. Inside the thin layer $y<2L_0$,  the grid is doubled in both the x and y-directions and the resolution is two times higher than in the other region at $t=0$. The highest refinement level is 11 in the simulations shown in this paper. A convergence study was carried out by repeating the simulation of Case~I with a higher resolution such that the highest refinement level is limited to 12. The numerical results in the higher resolution case are very similar to the results presented in the next section.

The staggered grids are applied in the numerical code. In 2D simulations, the magnetic field $b_x$ is located at the centers of the left and right edges of the rectangular grid cells, $b_y$ is located at the centers of the top and bottom edges, and the other variables are located at the centers of the rectangular grid cells. Therefore, if there are $n_x$ grid cells in the x-direction and $n_y$ grid cells in the y-direction for the other variables, the number of grid cells for $b_x$ are $n_x+1$ in the x-direction and $n_y$ in the y-direction, and the number of grid cells for $b_y$ are $n_y+1$ in the y-direction and $n_x$ in the x-direction. Assuming that the length for one rectangular grid cell is $dx$ in the x-direction, there will be a distance of $0.5dx$ between the grid location for $\rho(n_x, n_y)$ and the grid location for $b_x(n_x, n_y)$. 

The raw data calculated from the C code are usually transformed into the uniform IDL data, which are then used to plot all the figures presented in this work. In this process,  the original nonuniform adaptive staggered grids have been transformed to uniform unstaggered grids by using appropriate extrapolation. The two extra ghost layers mentioned above are not included in the IDL data. The values of all the variables which are located at the physical boundaries of the figures presented in this work are extrapolated by using data from the two ghost layers and the layers near the boundary inside the simulation box. Hence, the values of the corresponding variables at the boundary layers presented in all the figures here are close to the values in the ghost layers from the original raw data, but they are not exactly the same. The magnetic fields and plasma velocities at the physical bottom boundaries at four different times in Case~I\@ in the IDL data are presented in Figure~1. Since the magnetic field in the two ghost layers increases with time as shown in the above equation before $t=250$~s, the corresponding magnetic fields at the physical boundaries also increase with time before $t=250$~s. After $t=250$~s, the magnetic field does not change much at the bottom boundary for a long time. During the emergence process, we find that the current appears at the boundary inside $70L_0\leq x \leq 130L_0$. However, the strong magnetic diffusion in this region as mentioned above dissipates the current at the bottom. Hence, the current  density at the bottom boundary is still much smaller than the current density inside the main current sheet between the emerged and the background magnetic fields. One can choose different levels of uniform IDL output data. Since the level 11 IDL data are extremely large, all the figures in this work are plotted by using level 3 or level 4 uniform IDL data.  

\section{Numerical Results}\label{s:results}
\subsection{Jet morphology and dynamics in Case~I}\label{ss:jet:I}
As the magnetic flux emerges from the bottom, loop like magnetic structures are gradually formed at the bottom of the simulation domain. Magnetic reconnection between the background magnetic field and these loop structures leads to the formation of a current sheet at the left side of the loop structures. As more magnetic flux emerges, the loop structures become larger and the current sheet gradually becomes longer. The distributions of  the current density $J_z$, the temperature $T$, the logarithm of plasma number density $\mathrm{lg}\, {n}$, the velocity in the y-direction $v_y$, and the Mach number in Case~I at five different times after the emergence of magnetic fields stopped are presented in Figure~2. The long current sheet at $t=444$~s even extends up to $y=40$~km.  The maximum upward velocity in the y-direction ($v_y$) as shown on Figure~2(d) reaches $520$~km\,s$^{-1}$. The whole lifetime of the jet in Case~I is about 20 minutes. 

The upward reconnection outflows with high kinetic energy strongly collide with background magnetic field and plasma, and shocks form in this region as discussed below. The simplest form of a magnetic shock wave is the perpendicular one. In this case, the velocities of both the shock and the plasma are perpendicular to the magnetic field, which itself is unidirectional and parallel to the shock front. In a frame of reference moving with the shock, the Mach number for a perpendicular shock is defined as $Ma=\frac{v}{\sqrt{v_\mathrm{sound}^2+v_{A}^2}}$ \citep[e.g.,][]{2014masu.book.....P, 2015Sci...350.1238C}, where $v_\mathrm{sound}$ is the local sound speed and $v_{A}$ is the local Alfv\'en speed;  the plasma that is going through the shock undergoes a transition from a $Ma > 1$ to a $Ma < 1$ regime. Our simulations include multiple dynamical jump structures in which the magnetic field is generally not parallel to the plane of the discontinuity. For most of them,  it is very difficult to judge if they are shocks or not or measure the speed of these features. It is also very difficult for us to calculate the Mach numbers in the frames of reference moving with these jumps. In this work, the Mach number is calculated as $Ma=\frac{v}{\sqrt{v_\mathrm{sound}^2+v_{A}^2}}$, where $v$ is the absolute value of the local plasma velocity in the simulation reference frame. The highest Mach number as shown in Figure~2(e) can even reach $3.04$, which means the velocity of the reconnection outflows can be three times faster than the fast magneto-sonic wave.  The transition from a $Ma > 1$ to a $Ma < 1$ regime in Figure~2(e) can not be used to identify the possible shock regions.

Figure~3(a) shows a map of the divergence of the velocity field ($\nabla \cdot \mathbf{v}$) at $t=444$~s inside the box with red boundaries in the first panel of Figure~2(e). The abrupt jumps appear at the regions with high compression values along the three thick black dashed lines (SF1, SF2 and SF3). SF1, SF2 and SF3 do not show apparent motion around $t=444$~s, so we identify their velocities close to zero. In the following, assuming that the speed of these interfaces is zero greatly simplifies the analyses of MHD jump conditions. 

We analyze how different variables change along the direction perpendicular to these interfaces. From the MHD jump conditions \citep[]{2014masu.book.....P} (e.g.,  $B_{n1}=B_{n2}$, $\rho_1 v_{n1}=\rho_2 v_{n2}$, $\rho_1v_{n1}^2+p_1+\frac{B_{t1}^2}{2\mu_0}=\rho_2v_{n2}^2+p_2+\frac{B_{t2}^2}{2\mu_0}$, the subscript $t$ represents the component which is tangential to the shock front, the subscript $n$ represents the component which is normal to the shock front), we find that a slow mode shock appears at SF1, and it is very close to an intermediate shock. Fast mode shocks appear at SF2 and SF3. As shown in Figure~3(a), the thick arrow NL is perpendicular to SF3. This arrow starts at the point $x=106.5L_0, y=28.5L_0$ and ends at $x=115.0L_0,y=22.4L_0$. The length of this arrow is about $10.5L_0$. NL intersects with the shock front SF3 at $x=110L_0, y=25.8L_0$ when the length of NL reaches about $4.4L_0$. The distributions of different variables along NL are shown in Figure~3(b), (c), (d), (e) and (f). Figure~3(b) and (d) show that the plasma density ($\rho$), plasma pressure ($p$) and the magnetic field ($B_t$) which is parallel to SF3 strongly increase behind the shock front SF3. Figure~3(c) shows that the two components of the plasma velocity $v_t$ and $v_n$ both decrease behind the shock front SF3. The velocity component $v_n$ is nonzero both ahead and behind SF3, which confirms the plasma is going through the shock front. The magnetic field which is normal to SF3 ($B_n$) is almost unchanged ahead and behind SF3 as shown in Figure~3(d). The distributions of $\rho v_n$ and $\rho v_n^2+p+\frac{B_t^2}{2\mu_0}$ along NL are shown in Figure~3(e) and 3(f). For the most part we can ignore the changes of $\rho v_n^2+p+\frac{B_t^2}{2\mu_0}$ before and after the shock front, but  the changes of $\rho v_n$ are significant. Therefore, the jump conditions of $B_{n1}=B_{n2}$ and $\rho_1v_{n1}^2+p_1+\frac{B_{t1}^2}{2\mu_0}=\rho_2v_{n2}^2+p_2+\frac{B_{t2}^2}{2\mu_0}$ are almost satisfied, but  $\rho_1 v_{n1}=\rho_2 v_{n2}$ is not satisfied well. One should keep in mind that the theoretical analyses are usually idealized and all the physical variables in the MHD jump conditions are assumed to be uniform at each side of the shock front and time independent, but all the variables change with location and time in the numerical simulations. Therefore, it is not surprising that there are some deviations between the analytical and numerical results. We can still conclude that a shock appears at SF3. Since the parallel component $B_t$ sharply increases from zero to high values behind the shock front, the fast mode shock at SF3 is  similar to a switch-on shock. The methods for analyzing the jump conditions for SF1 and SF2 are the same as just described. The jump condition about $\rho_1 v_{n1}=\rho_2 v_{n2}$ for SF2 is satisfied better than SF3. 

The chromospheric jet simulations by \cite{2013ApJ...777...16Y} showed slow mode shocks propagating along the jet, but no fast mode shocks formed because plasma $\beta$ is much higher in their simulations. The strong fast mode shocks at SF2 and SF3 in our simulations drive part of the outgoing plasma upward along the open lines, thus creating an inverse Y-shape structure \citep{1995Natur.375...42Y, 1996PASJ...48..353Y}. Figure~2(c) shows that the high density plasma from the bottom is ejected outward along the jet. Figure~2(b) shows that  the plasma in the jet is strongly heated by the fast mode shocks. The temperature in the jet region is even higher than in the long main current sheet at the bottom. The average temperature in the jet is around $3.5$~MK\@. At SF1 where the slow mode shock appears, the plasma is only weakly heated. Figure~11 from \cite{2016ApJ...822...18N} shows an analysis of the shock structures in the outflow region of the main current sheet. However, their shock structures differ from ours. They found wedge like shock fronts, where the upper part is roughly horizontal and nearly perpendicular to the field lines in the post-shock region and resembles a switch-off slow mode shock; the fast mode shock appears at the lower part and is almost vertical branch of the wedge. The slow mode shock in the upper part is directly related to the hot jet in the simulations by \cite{2016ApJ...822...18N}. However, the fast mode shocks heat the plasma in the jet in this case in our work. Plasma $\beta$ is an important factor in causing these differences.     

After $t=444$~s, as magnetic reconnection between the emerged and the background magnetic fields continues, the upward outflow region gradually becomes more chaotic. As shown in Figure~2(a), many current sheet fragments appear above the cusp shape in the jet region at $t=561$~s, $t=678$~s and $t=759$~s. These fragments become unstable and magnetic islands first appear in this region. This phenomenon of magnetic islands appearing in the jet region (in addition to the main current sheet region) are not shown in previous simulations of jets. The plasma in this complicated area in the jet region is strongly heated as shown in Figure~2(b). The maximum local temperature in the small magnetic islands can reach around $8$~MK\@. As discussed in section 2, the heating term $\mathcal{H}$ is small and contributes very little heat in this region. From Figure 2(a), one can see that the current density in these small islands is also high. Hence, Joule heating plays an important role in heating the plasma inside these islands. On the other hand, the reconnection outflows from the nearby reconnection X-points are very strong in the low $\beta$ plasma. As the high velocity plasmas flow into these islands, the kinetic energy can be converted to thermal energy by strong compressions at the slow and fast mode shocks inside the islands. The multiple slow mode and fast mode shocks inside the magnetic islands have been studied and analyzed in detailed in several previous papers \citep[e.g.,][]{2011PhPl...18b2105Z, 2015ApJ...799...79N, 2016ApJ...832..195N}. Subsequently, the magnetic islands gradually appear in the main current sheet of the jet leg. Since the main reconnection X-point is located at a very low height which is close to the bottom boundary at the beginning, almost all of the small magnetic islands in the main current sheet move upward. As the main X-point gradually rises to higher heights, both upward and downward propagating magnetic islands appear during the later phase. A flare-like bright loop near the bottom boundary then forms. Magnetic islands with different velocities can coalescence with each other to form bigger islands. After the magnetic islands appear in the main current sheet, the shock structures in the outflow region of the main current sheet gradually break down as shown in Figure~2(e). 

In order to compare with the observations, we calculate the temperature and density dependent emission count rate. Figure~4(a) presents the emission count rate $\mathrm{ECR}=\int n^2f(T)dl$~DN\,s$^{-1}$\,pixel$^{-1}$, where $f(T)$ is the AIA 335\,\AA\ response function \citep{2012SoPh..275...17L} from the Chianti package \citep{2015A&A...582A..56D},  $n$ is the number density, and $dl$ is the line element along the line of sight. This method for calculating the emission count rate from numerical simulations is the same as in previous papers \citep[e.g.,][]{2014ApJ...796L..29G}. Since the variables do not vary in $z$-direction, we can get $\mathrm{ECR}=n^2f(T)z_h$, where $z_h=10^7$~m is the assumed width of the jet along the $z$-direction. 

The emission count rate of the thin current sheet and jet is strong at $t=444$~s. As the outflow region becomes chaotic, the bright blobs shown in the second, third, and the fourth panels appear and move upward along the jet. These features are similar to the blobs in the jet observed by \citet{2014A&A...567A..11Z} and \citet{2016SoPh..291..859Z}.  As we zoom into the small box surrounded by the white dotted boundaries in the third panel at $t=678$~s in Figure~4(a), there are two bright blobs which have strong emissions in this domain. We find that there is no O-point inside the bright blob at the lower position and there are no obvious current sheet fragments and reconnection X-points near it. The bright blob at the higher position is a magnetic island containing an O-point, and there are reconnection X-points nearby. Therefore, we can conclude that the bright blob at the lower position is not a magnetic island but a blob with high plasma density and relatively high temperature [see Figure~4(b), (c), and (d)], while the bright blob at the higher position is indeed a magnetic island produced by plasmoid instability. Many bright blobs in the fourth panel at $t=759$~s in Figure~4(a) are also magnetic islands. The detailed information inside the two boxes surrounded by the black dotted boundaries in the fourth panel in Figure~2(a) at $t=759$~s is presented separately in Figure~5 and Figure~6. Figure~5 corresponds to the bottom box and  shows the main current sheet region. The multiple magnetic islands of different sizes are all dense and hot, which can both be identified as blobs in the AIA hot channels (335\,\AA\ and 211\,\AA ) and the AIA cool channel (304\,\AA). As shown in Figure~5(b) and 5(c), the highest temperature around $2.5$~MK is located in the center of the big magnetic island at around $y=10L_0$, the temperature in the center of the other islands is also much higher than their boundaries and the highest mass density is also in the center. The response function $f(T)$ in AIA channel 335\,\AA\ covers a wide range around $4\times10^4\sim1.6\times10^7$~K and has three high value peaks at around $2\times10^5$~K, $8\times10^5$~K and $2.8\times10^6$~K; $f(T)$ in AIA channel 211\,\AA\ covers a range around $6\times10^4\sim6\times10^6$~K and has a high value peak at $2.5\times10^5$~K and a low value peak at $1.7\times10^6$~K;  $f(T)$ in AIA channel 304\,\AA\ covers a range around $4\times10^4\sim2.5\times10^6$~K and has a high value peak at $8\times10^4$~K and an extremely low value peak at $1.6\times10^6$~K. Therefore, the high temperature and high density blob at around $y=10L_0$ is very bright in the center in AIA channel 335\,\AA. The temperature around the boundary of this blob is about $1.6$~MK which is almost corresponding to one of the peak values in both AIA chancel 211\,\AA\ and 304\,\AA. Therefore, the boundary of this blob is brighter than its center in both channel 211\,\AA\ and 304\,\AA. As shown in Figure~6, numerous magnetic islands are generated in the chaotic jet region in the upper box. Only the magnetic islands with relatively high number density located in the left part of this box can be identified in the AIA channels (335\,\AA, 211\,\AA\ and 304\,\AA). The magnetic islands with much higher temperature but lower number density near the right boundary are not identified in the AIA channels. The emission count rate strongly depends on the number density, which is also apparent when comparing Figure~4(b), (c) and (d). There is a major magnetic island toward the bottom right of this domain, and the temperature at the center of this island exceeds $5$~MK. However, the mass density inside this major island is lower than that inside the two bright blobs shown in Figure~4(b), and this major island is dark in the AIA 335\,\AA\ channel. The above phenomenon that the bright blobs in the jet can be identified in different AIA channels with different temperature ranges is similar to the observational results by \citet{2014A&A...567A..11Z} and \citet{2016SoPh..291..859Z}.

As shown in Figures~2, 4, 5 and 6, the median temperature of the blobs that can be identified in the AIA channels is around $3$~MK\@. By observing and tracing the bright blobs, we find the maximum diameter of the blobs is around $6$~Mm, the maximum lifetime is around  $120$~s, and the  maximum velocity is around $200$~km\,s$^{-1}$. These characteristics of the simulated blobs are very similar to the blobs observed by \emph{SDO}/AIA and the Extreme-Ultraviolet Imager (EUVI) on \emph{STEREO} \citep{2014A&A...567A..11Z, 2016SoPh..291..859Z}.

\subsection{Jet morphology and dynamics in Case~II}\label{ss:jet:II}

The initial magnetic field in Case~II is half of that in Case~I, so the corresponding plasma $\beta$ in Case~II\@ is four times larger than that in Case~I\@. As shown in Figure~7, the jet lifetime in Case~II is about 40 minutes which is two times longer than in Case~I\@. Magnetic islands ejected along the jet were present in Case~I and showed strong emissions in three AIA channels. In contrast, there are significantly fewer islands in the jet region in Case~II\@ as shown in Figure~7(a). The islands that do form are very small and do not show obvious emissions in AIA channels. However, numerous chaotic current sheet fragments appear above the cusp in the jet in Case~II. These chaotic current sheet fragments can extend to the top of the jet, as shown in the second and the third panels  of Figure~7(a). Figure~7(b) shows that the maximum temperature in the jet region is about $1.9$~MK, which is four times lower than in Case I. Three obvious hot and high density blobs appear in the second and third panels of Figure~7(b) and (c). The maximum velocity in $y$-direction as shown in Figure~7(d) is $262$~km\,s$^{-1}$, which is about two times lower than in Case~I. Figure~7(e) shows that the structures with magneto-sonic Mach number higher than 1 appear not only in the outflow region of the main current sheet, but also at several places along the jet to the top. The high magneto-sonic Mach number structures also appear along the jet in Case~I at early times, but are then blocked by the magnetic islands that were generated above the cusp as shown in Figure~2(e)

Figure~8(a) shows the predicted emission count rate in the AIA 211\,\AA\, channel at five different times. The bright blobs move upward along the jet. The diameter of these blobs is around $8$~Mm. The blob at the lowest position appears first, followed by the blob in the middle and the blob at the top. The blobs move up slowly with an average velocity around $40$~km\,s$^{-1}$ soon after their formation. After about 3 minutes, these bright blobs move toward the bottom left direction and finally disappear. Figures~8(b), (c), (d), (e) and (f)  show the emission count rate, temperature, the logarithm of plasma number density, Mach number and vorticity in the box with white dotted boundaries in the second panel at $t=1085$~s in Figure~8(a). All these variables have abrupt changes at multiple interfaces. In Figure~8(e), F1, F2, F3, F4, F5 and F6 are six detected interfaces where the magneto-sonic Mach number changes abruptly. The distributions of three variables along the two black dashed lines at $y=60L_0$ and $y=66.3L_0$ in Figure~8(e) are presented in Figure~9. The two black dashed lines pass through the interface F3 at $x=116.4L_0$ and F4 at $x=122.5L_0$ separately. In Figure~9(a), the black solid line represents the magnetic fields which are parallel to F3 and the red dotted line represents the magnetic fields which are parallel to F4. Figure~9(b) and (c) show the distributions of the temperature and plasma density along x-direction. All these variables have drastic changes along x-direction. We can not find a good method to measure the speed of these interfaces. Therefore, it is impossible for us to tell if the MHD jump conditions can be satisfied or not in the frames of reference moving with these interfaces. We can not make conclusions that the shocks appear there.              

 Figure~8(f) shows that vortex-like structures appear in these blobs. Therefore, the Kelvin-Helmholtz instability \citep[e.g.,][]{1999JPlPh..61....1K,2016ApJ...824...60T} is triggered by the strong shear flows between the high velocity jet and the roughly stationary ambient plasma. Since the high Mach number indicates that the Alfv\'en speed is much slower than the local plasma velocity, the Kelvin-Helmholtz instability is not suppressed by the parallel magnetic fields in the high Mach number regions. However, the high Mach number regions in the jet do not persevere for a long time. As shown in Figure.~7, they eventually break down and disappear after secondary islands appear in the main current sheet. The Kelvin-Helmholtz instability is then suppressed and the bright blobs eventually move downward and disappear. Simultaneously, the hot and dense magnetic islands in the main current sheet region are gradually growing larger. 

\subsection{Jet morphology and dynamics in Case~III}\label{ss:jet:III}

The initial magnetic field in Case~III is $b_0=0.00075$~T, which is one quarter of that in Case~I\@. The corresponding initial plasma $\beta$ in Case~III is 16 times larger than in Case~I\@. Larger plasma $\beta$ corresponds to lower magnetic field or higher plasma pressure. In the three cases in our simulations, the initial plasma density and pressure distributions are the same and only the initial magnetic field decreases linearly. The weaker magnetic field leads to less magnetic energy being converted to thermal energy and kinetic energy during reconnection in our cases. Therefore,  the plasma in the reconnection outflow region will get less kinetic energy and thermal energy. For plasma with similar mass density, the reconnection outflows will be heated to lower temperatures and reach lower velocities. Table~1 shows that the maximum velocity of the jet almost linearly decreases for the weaker initial magnetic fields. Since the length scale is the same, the lower velocity leads to a longer duration. The magneto-sonic Mach number $Ma=\frac{v}{\sqrt{v_\mathrm{sound}^2+v_{A}^2}}$ relates the local velocity, plasma density and magnetic field, and will change with different magnetic field. From the three cases, we can only conclude that the weaker magnetic field results in the lower magneto-sonic Mach number and weaker shock waves, which then leads to the weaker compression and lower temperature increases behind the shock fronts. As shown in Figure~10, the maximum temperature is only about $0.9$~MK, the maximum velocity in the $y$-direction is about $90$~km\,s$^{-1}$, and the jet lifetime is about 80 minutes in Case III\@. The maximum Mach number is only $0.9$, and compression by the reconnection out flow is much weaker in Case~III than in Case~I and Case~II\@. The resulting jet has a much lower velocity and the maximum temperature increase is only around $0.1$~MK\@.  As shown in Figure~10(c), there remains some high density plasma which can be ejected up along the jet, and a clear vortex appears in the reconnection outflow region above the cusp. However, no bright blobs are ejected out along the jet. The reconnection outflow region is less chaotic in Case~III than in Case~I or Case~II\@. Fewer magnetic islands are generated in the main current sheet near the bottom in Case III than in Case~I or Case~II\@.

The length of the main current sheet is around $40L_0$ before secondary islands appear as shown in Figure ~2(a),  7(a) and 10(a). The corresponding temperature inside the main current sheet is around $2.8$~MK in Case~I\@, $1.2$~MK in Case~II\@ and $0.8$~MK in Case~III\@, and the plasma density is around $6.618\times10^{-11}$~kg\,m$^{-3}$ in all the three cases. We use the initial magnetic field $b_0$ to calculate the Alfv\'en speed around the main current sheet. Then,  the physical Lundquist number $S=Lv_A/\eta$ in the main current sheet before secondary islands appear is calculated to be $8.824\times10^5$ in Case~I\@, $1.209\times10^5$ in Case~II\@ and $0.328\times10^5$ in Case~III. We have run an additional case which is identical to Case~III, but with a much smaller constant magnetic diffusivity ($\eta=3.5\times10^6$~m$^2$\,s$^{-1}$). However, the numerical results in this additional case are similar to those in Case~III\@. The different numerical results shown in Case~I, Case~II, and Case~III above are therefore mostly affected by plasma $\beta·$ rather than the Lundquist number. 

\section{Summary and discussions}\label{s:discussion}

The mechanisms responsible for the formation of bright blobs in coronal jets remain unknown. In this work, we study jet formation triggered by magnetic reconnection between newly emerged magnetic flux and background magnetic fields using 2D numerical simulations. We include optically thin radiative cooling and heat conduction, and use a high Lundquist number. We present three cases with different magnetic fields and plasma $\beta$. The numerical results and conclusions are summarized as below:

(1) The magnetic reconnection outflows can strongly compress the background magnetic fields and plasma to form fast mode shocks, which then drive part of the outgoing plasma upward along open field lines to create an inverse Y-shaped jet. The outflow velocities, temperature, the number and properties of magnetic islands in the jet strongly depend on plasma $\beta$. For lower plasma $\beta$ with stronger magnetic fields or lower plasma density, the lifetime of the jet is shorter, the Mach numbers of the fast mode shocks in the reconnection outflow regions are higher, the jet and magnetic islands are formed with higher velocities and temperatures, the current sheet fragments are more chaotic in the jet, and more magnetic islands are generated. In Case~I in our simulation, the lifetime of the jet is around 20 minutes, the maximum upward velocity is around $520$~km\,s$^{-1}$ and the maximum temperature is around $8$~MK\@. These characteristics are consistent with the observed properties of coronal jets.   

(2) For Case~I with low plasma $\beta$, magnetic islands are generated not just in the main current sheet region below the jet, but also in the jet itself.  We show for the first time that magnetic islands are ejected out along the jet in the numerical simulations. These islands can be identified in the AIA hot channels (335\,\AA\, and 211\,\AA\,) and the AIA cool channel (304\,\AA\,). The median temperature is around $3$~MK, the maximum diameter is around $6$~Mm, the maximum lifetime is around $120$~s and the maximum velocity is around $200$~km\,s$^{-1}$.  The characteristics of these magnetic islands are similar to the observed bright and dense blobs in the EUV jets by multiple AIA channels. Because plasma $\beta$ in real coronal jets can be as low as or even lower than in Case~I, some of the bright blobs observed in jets are likely to correspond to magnetic islands generated by the plasmoid instability in chaotic current sheet fragments in the jet.

(3) High temperature and high density vortex-like blobs form along the jet in Case~II with moderately high plasma $\beta$. These vortex structures are above the high magneto-sonic Mach number regions and likely formed in the Kelvin-Helmholtz instability process. These bright vortex-like blobs can also be identified in the AIA channels, and they move up along the jet for about three minutes in Case~II\@. However, they eventually move downward and disappear after the high magneto-sonic Mach number regions are broken down by the gradually appearing magnetic islands in the main current sheet.

Our numerical results provide two possible mechanisms to explain the generation of bright blobs in the EUV jets: the plasmoid instability operating in the chaotic region above the cusp in the jet, and Kelvin-Helmholtz instability along the jet. Our simulations show that plasma $\beta$ is a very important parameter in determining the temperature, velocity, and lifetimes of the jet and blobs. In an environment with a relatively high Lundquist number and low plasma $\beta$, magnetic islands are usually easily formed. Many previous observations did not observe bright blobs inside jets. The observability of blobs in jets depend on both the space and time resolutions and viewing angle.  However, there could be some other physical mechanisms that hinder the formation of the blobs, e.g. the guide field in the jet and 3D magnetic reconnection structures can also possibly affect the formation of jets and associated blobs. Therefore, we still need further theoretical and observational investigations into how blobs are generated and why no apparent blobs form in some jets.
 
\section*{ACKNOWLEDGMENTS}
The authors are grateful to the referee for the valuable comments and suggestions. This research is supported by NSFC Grants 11573064, 11203069, 11273055, 11333007, 11303101 and 11403100; the Western Light of Chinese Academy of Sciences 2014; the Youth Innovation Promotion Association CAS 2017; Key Laboratory of Solar Activity grant KLSA201404; Program 973 grant 2013CBA01503; the Youth Fund of Jiangsu Province BK20161618; CAS grant XDB09040202; and the Special Program for Applied Research on Super Computation of the NSFC-Guangdong Joint Fund (nsfc2015-460, nsfc2015-463, the second phase). N.A.M.\ acknowledges support from NSF SHINE grants AGS-1156076 and AGS-1358432, and DOE grant DE-SC0016363 through the NSF/DOE Partnership in Basic Plasma Science and Engineering. We have used the NIRVANA code v3.6 and v3.8 developed by Udo Ziegler at the Leibniz-Institut f\"ur Astrophysik  Potsdam. The authors gratefully acknowledge the computing time granted by the Yunnan Astronomical Observatories and the National Supercomputer Center in Guangzhou, and provided on the facilities at the Supercomputing Platform, as well as the help from all faculties of the Platform.

\clearpage

\begin{table*}
 \caption{Jet parameters for Cases I, II and III\@. Here, $b_0$ is the initial magnetic field, $\beta_{0}$ is the initial plasma $\beta$, $t_\mathrm{life}$ is the jet lifetime, $T_\mathrm{max}$ is the maximum temperature in the jet, $v_\mathrm{ymax}$ is the maximum velocity in the $y$-direction of the jets, and $Ma_\mathrm{max}$ is the maximum Mach number in the jets.}
\label{Parameter}
  \begin{tabular}{lcccccc}
   \hline
                    &$b_0$(T)     &      $\beta_0$            &$ t_\mathrm{life}$ (min) &  $T_\mathrm{max}$(MK)&  $v_\mathrm{ymax}$(km/s) &$Ma_\mathrm{max}$      \\
    \hline
    Case I     &$0.003$      &   $0.0393 -0.3083$ &  $20$                               &  $8 $                                    & $520$                                      & $3.04$\\ 
   \hline
    Case II    &$0.0015$    &   $0.1574-1.5216$  &  $40 $                              &  $2$                                     & $260$                                       & $2.02$\\ 
    \hline
    Case III  &$0.00075$  &  $0.6296 - 4.9349$ & $80$                                & $0.9$                                   & $90$                                         & $0.9$\\ 
    \hline   
  \end{tabular}   
\end{table*}

\begin{figure}
   \centerline{\includegraphics[width=0.87\textwidth, clip=]{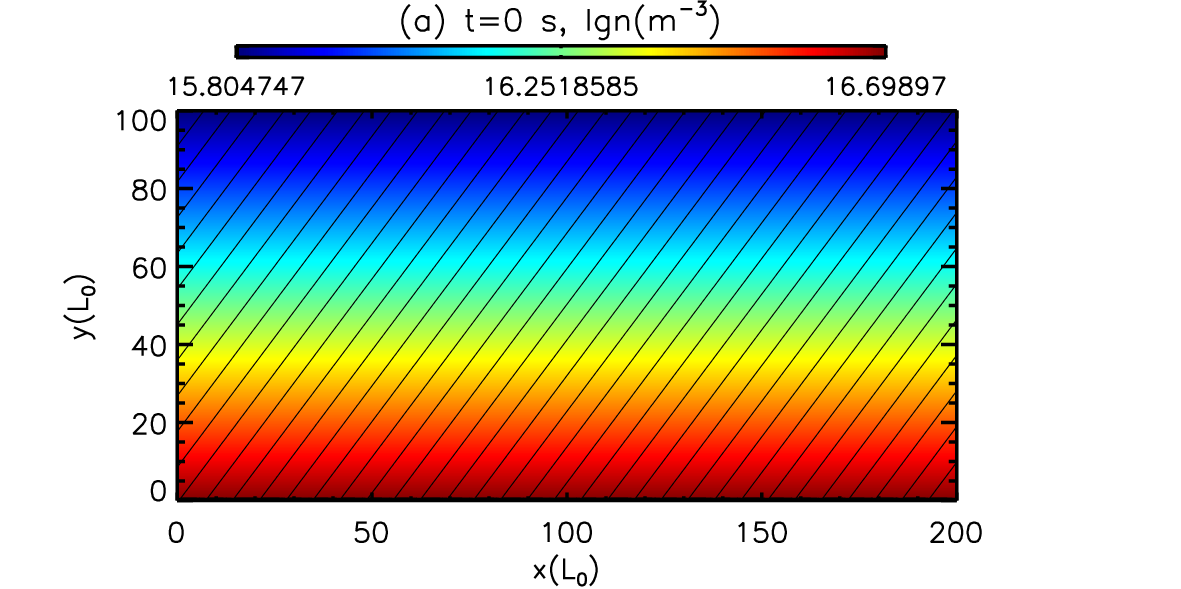}}
    \centerline{\includegraphics[width=0.38\textwidth, clip=]{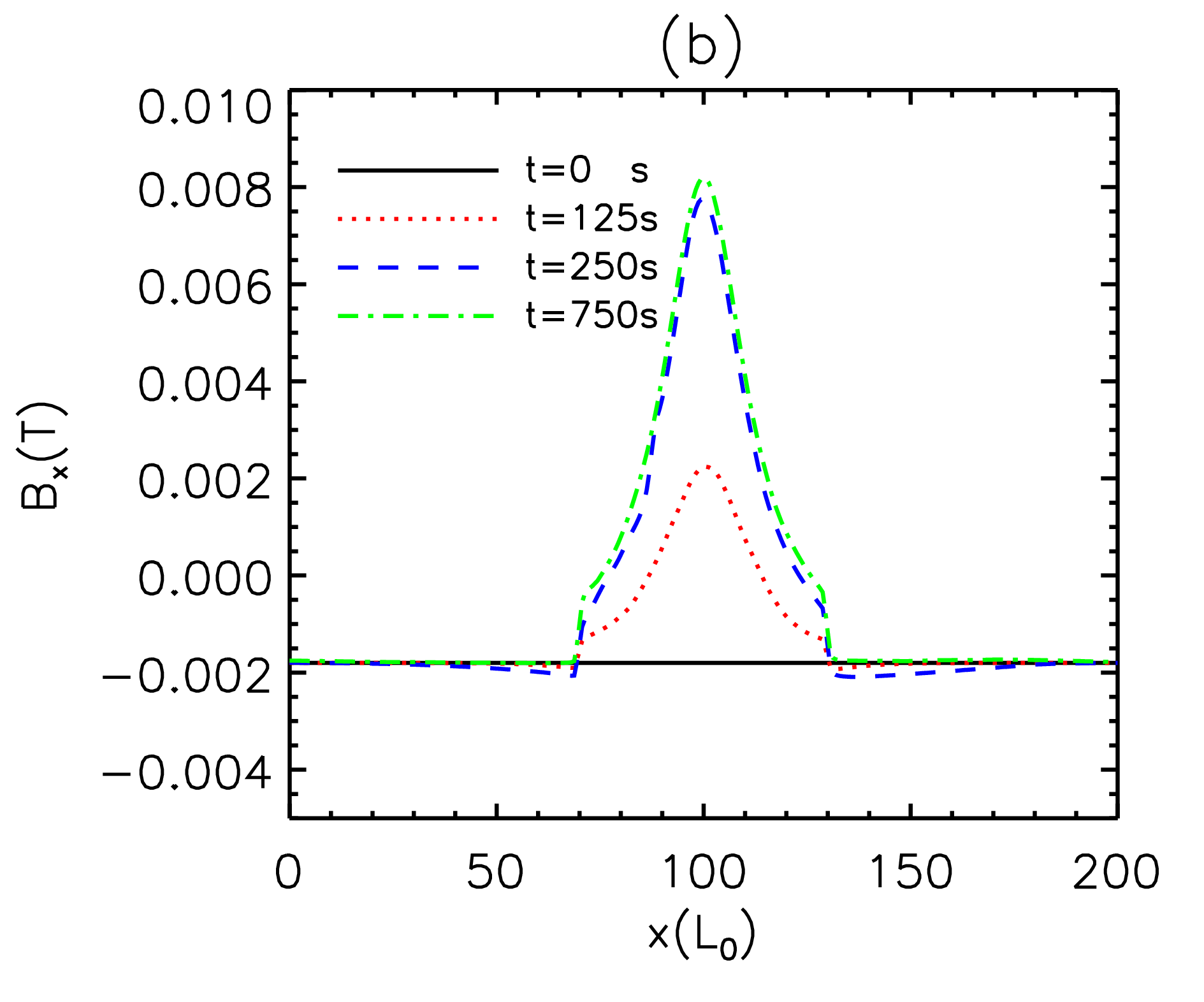}
                       \includegraphics[width=0.38\textwidth, clip=]{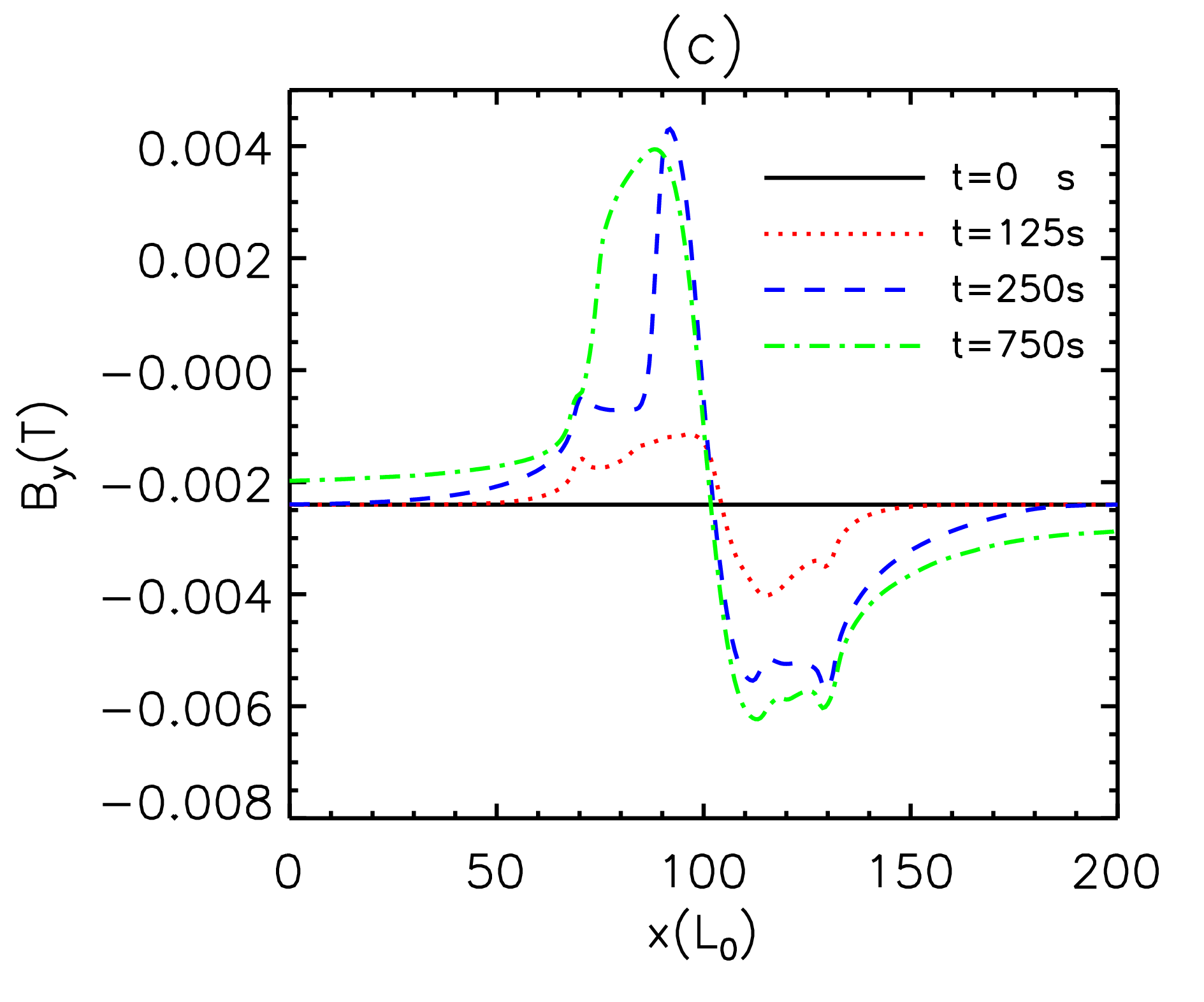}}
    \centerline{\includegraphics[width=0.38\textwidth, clip=]{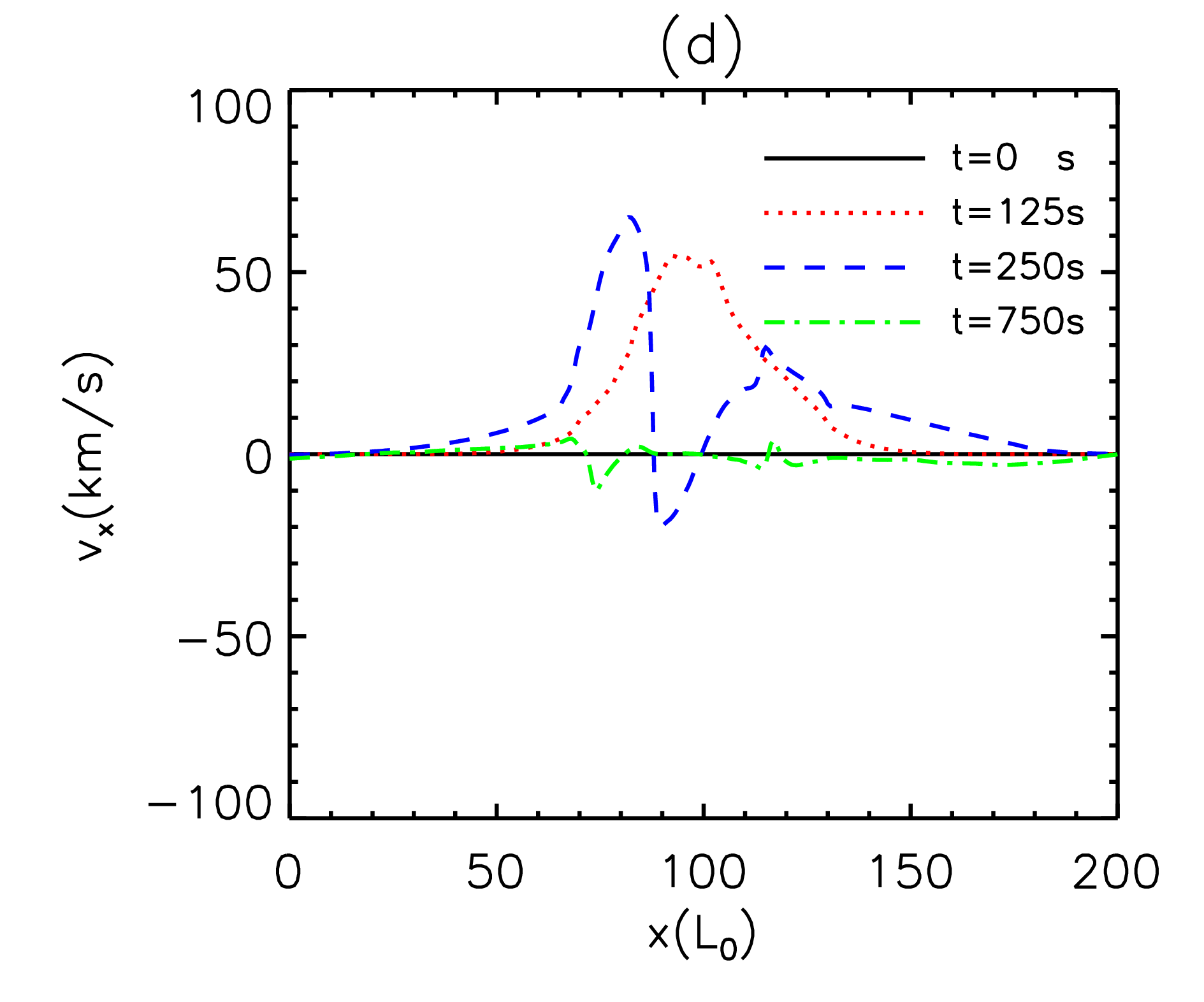}
                         \includegraphics[width=0.38\textwidth, clip=]{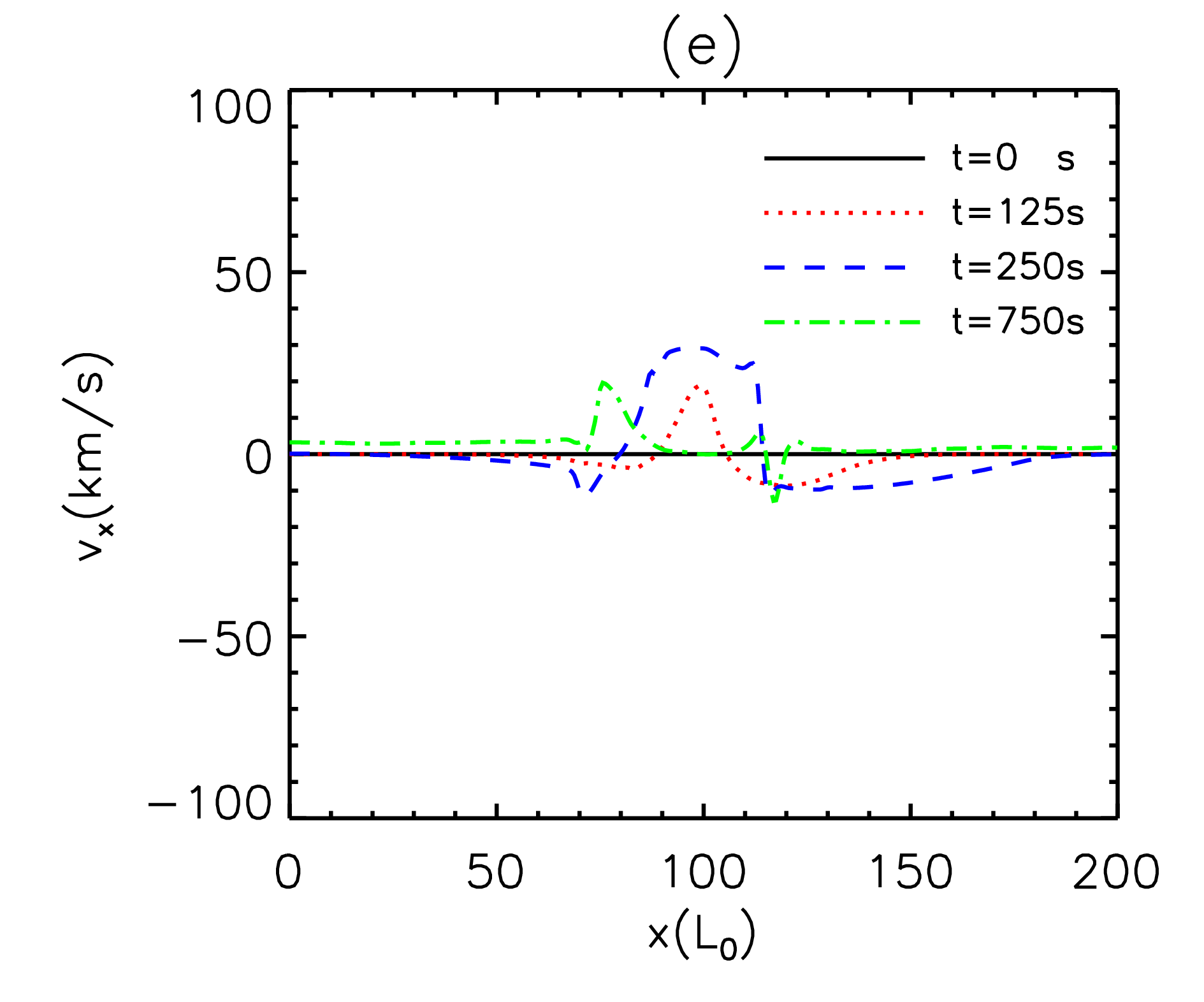}}
   \caption{ (a) The initial conditions for Case~I\@.  The color contour represents the logarithm of the initial number density and the black solid lines represent the initial magnetic field. The magnetic field (b) Bx,  (c) By,  and velocity (d) $v_x$ and (e) $v_y$  at $y=0$ are distributed along x direction at four different times.}
   \label{fig.1}
 \end{figure}

\begin{figure*}
      \centerline{\includegraphics[width=0.95\textwidth, clip=]{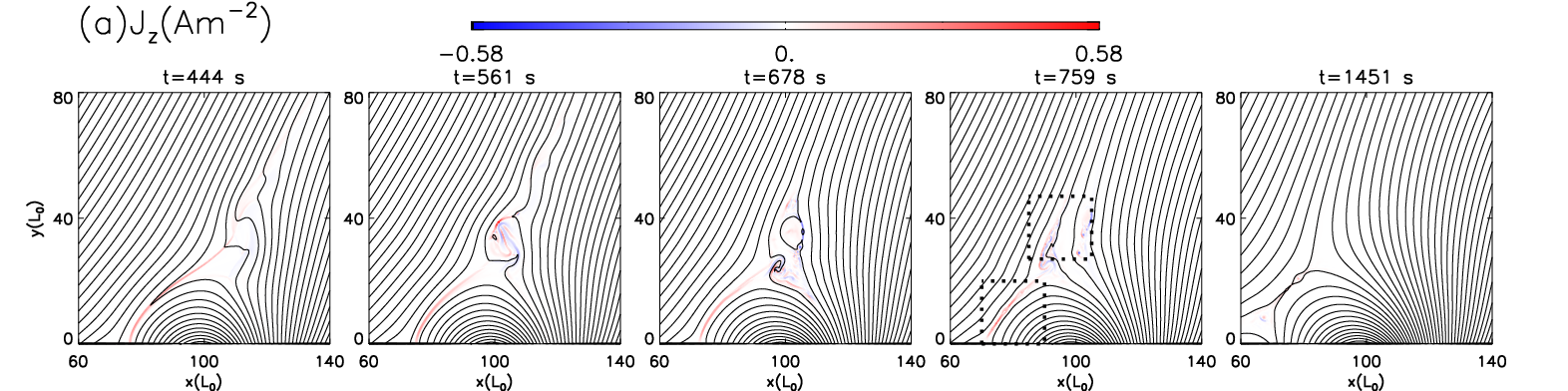}}
      \centerline{\includegraphics[width=0.95\textwidth, clip=]{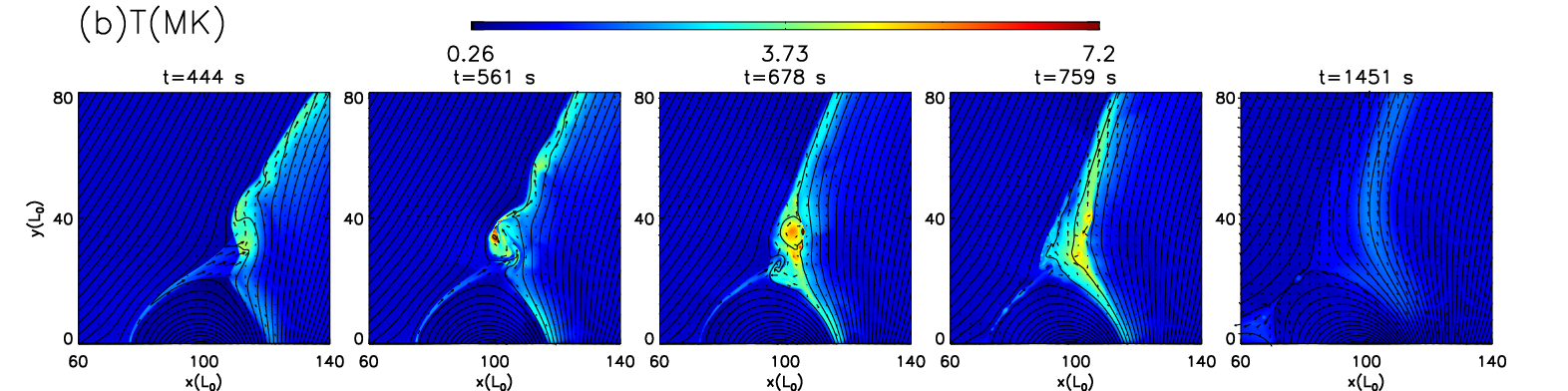}}
      \centerline{\includegraphics[width=0.95\textwidth, clip=]{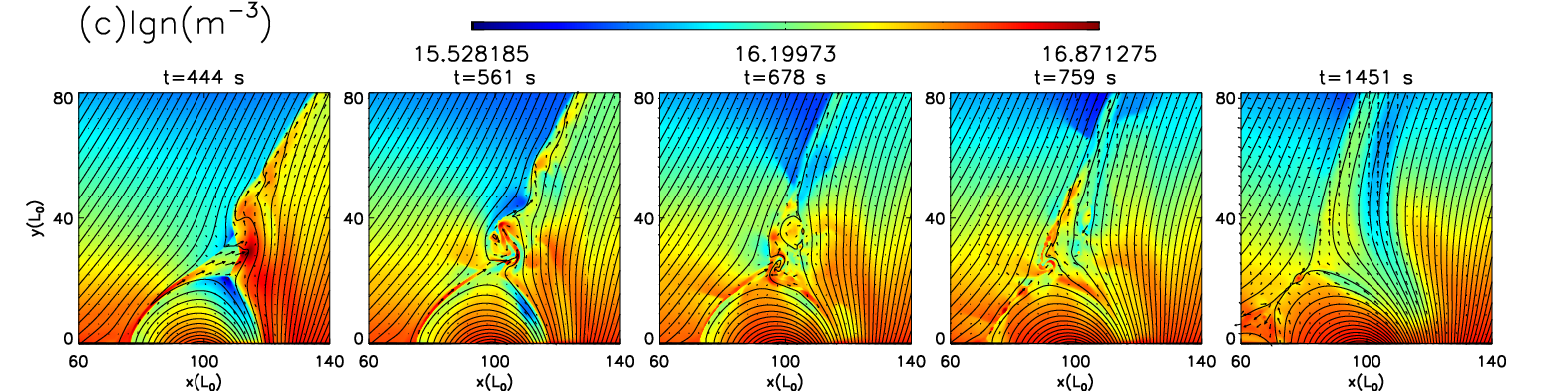}}
      \centerline{\includegraphics[width=0.95\textwidth, clip=]{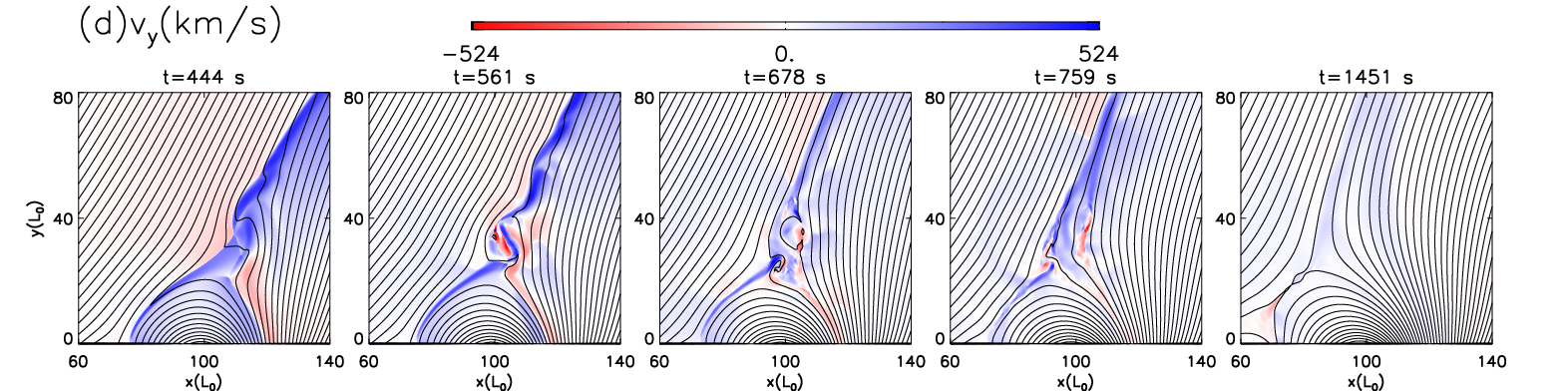}}
      \centerline{\includegraphics[width=0.95\textwidth, clip=]{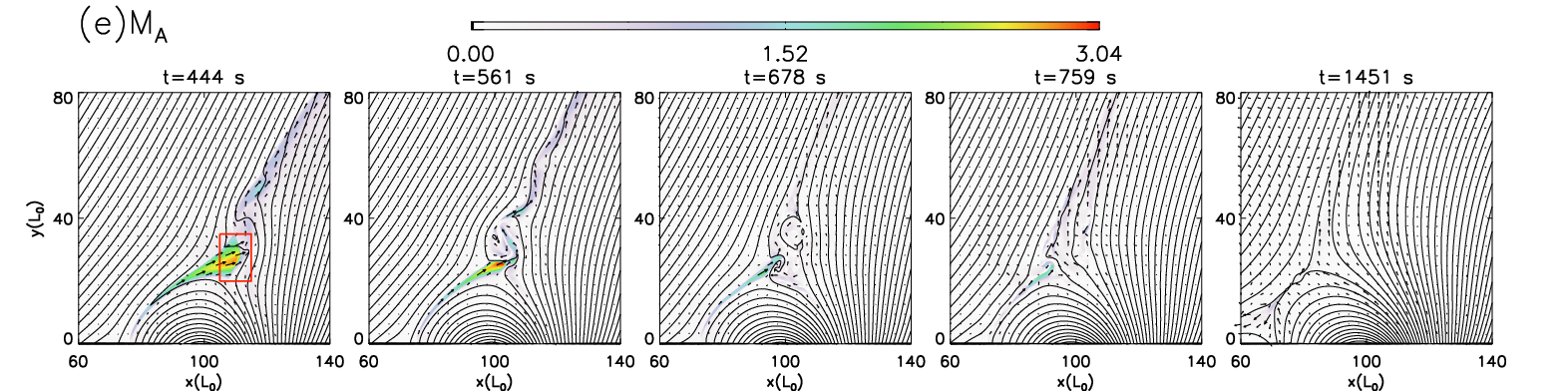}}
    \caption{The distributions of different variables at five different times in Case~I, including (a) current density, $J_z$; (b)temperature, $T$; (c) logarithm of plasma number density, $\mathrm{lg}\, n$; (d) velocity in the $y$-direction $v_y$; and (e) Mach number $M_a$. The black solid lines represent the magnetic fields and the black arrows represent the velocity vector in each panels.}
\label{fig.2}
\end{figure*}

\begin{figure}
  \centerline{\includegraphics[width=0.3\textwidth, clip=]{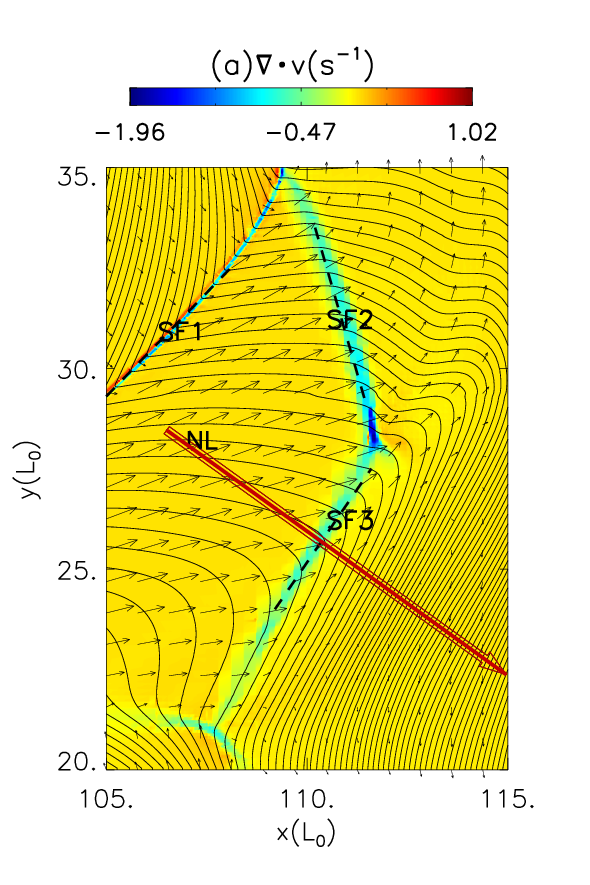}
                       \includegraphics[width=0.3\textwidth, clip=]{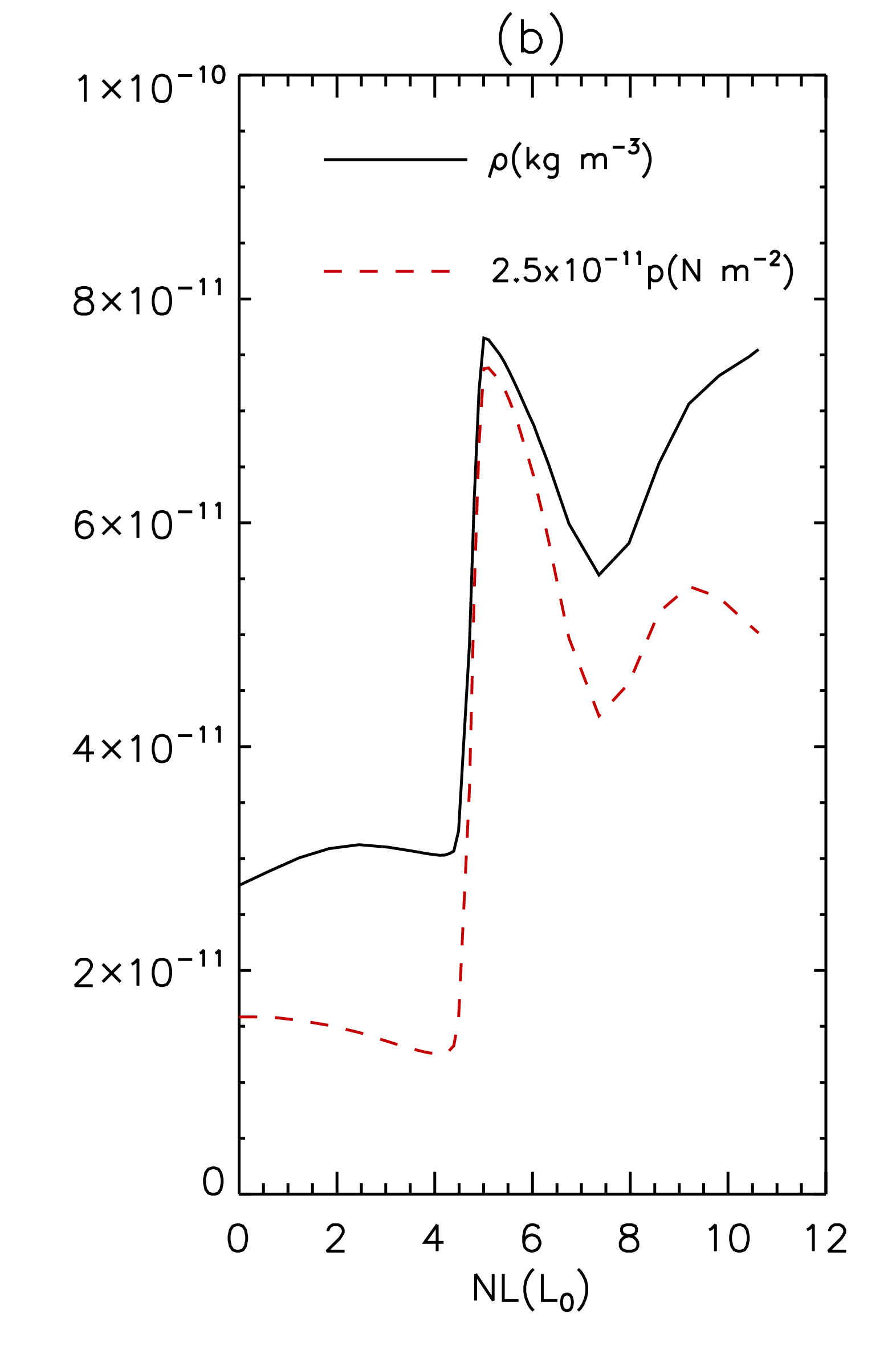}
                        \includegraphics[width=0.3\textwidth, clip=]{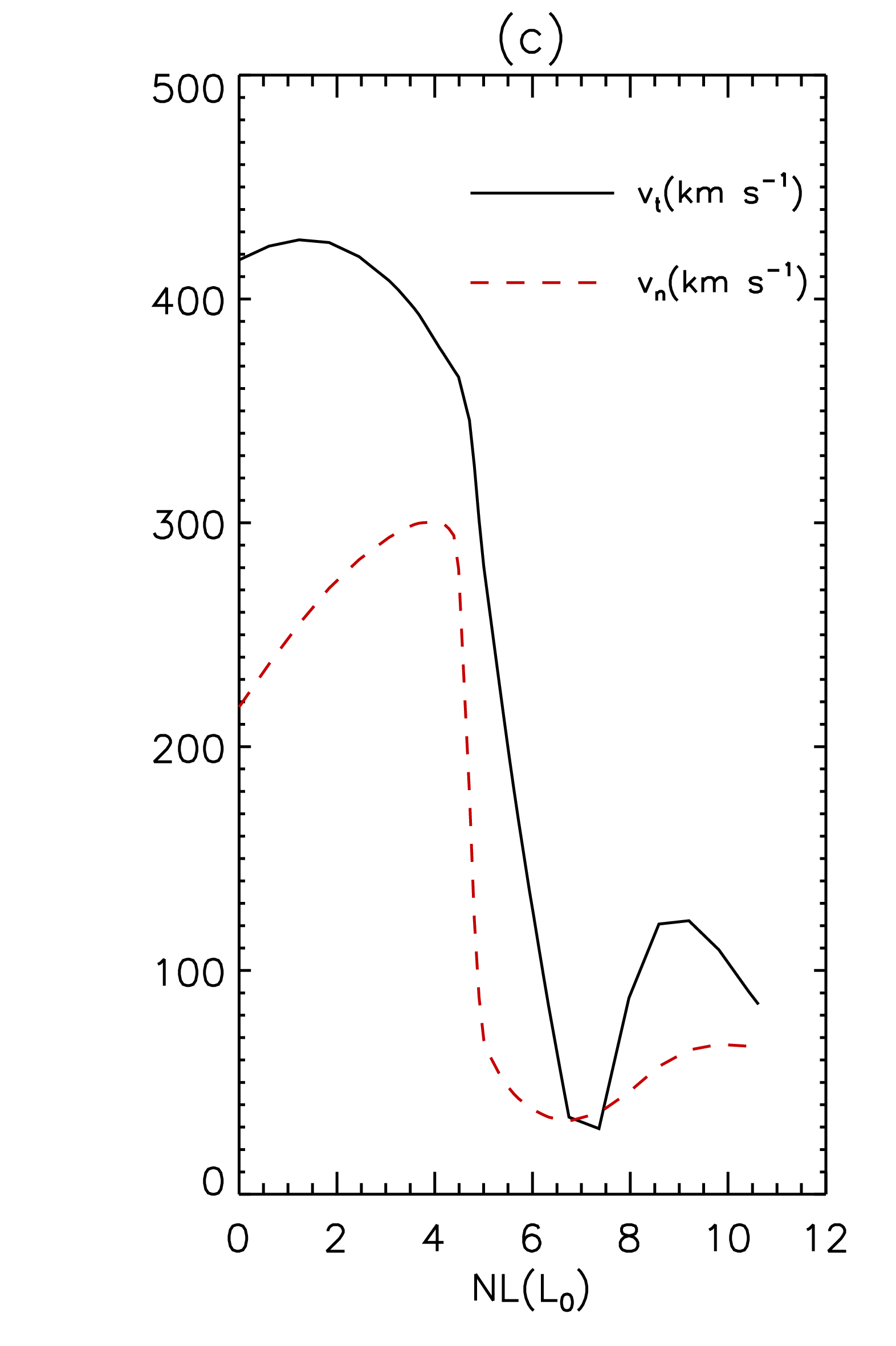}}
   \centerline{\includegraphics[width=0.3\textwidth, clip=]{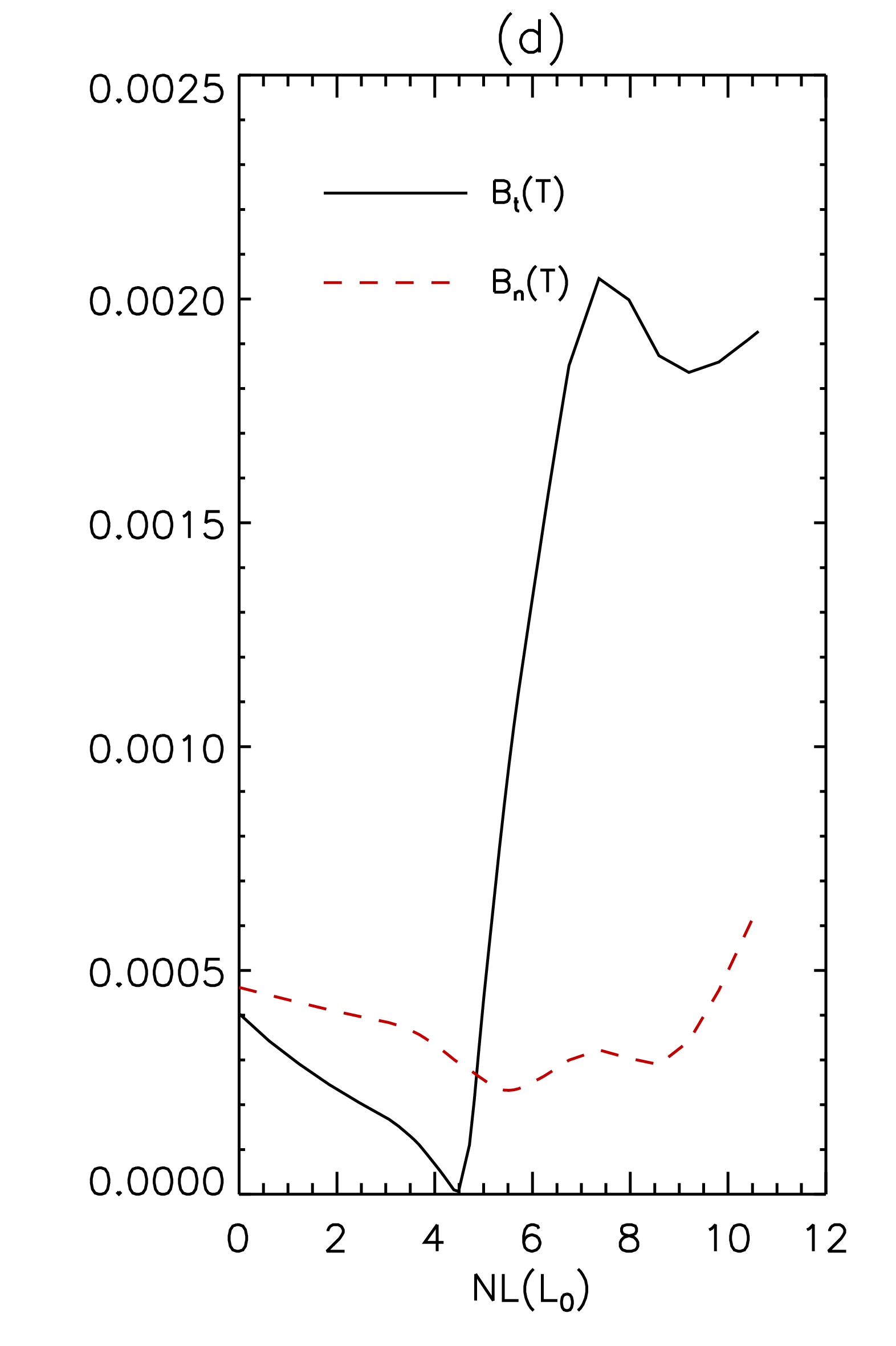}
                       \includegraphics[width=0.3\textwidth, clip=]{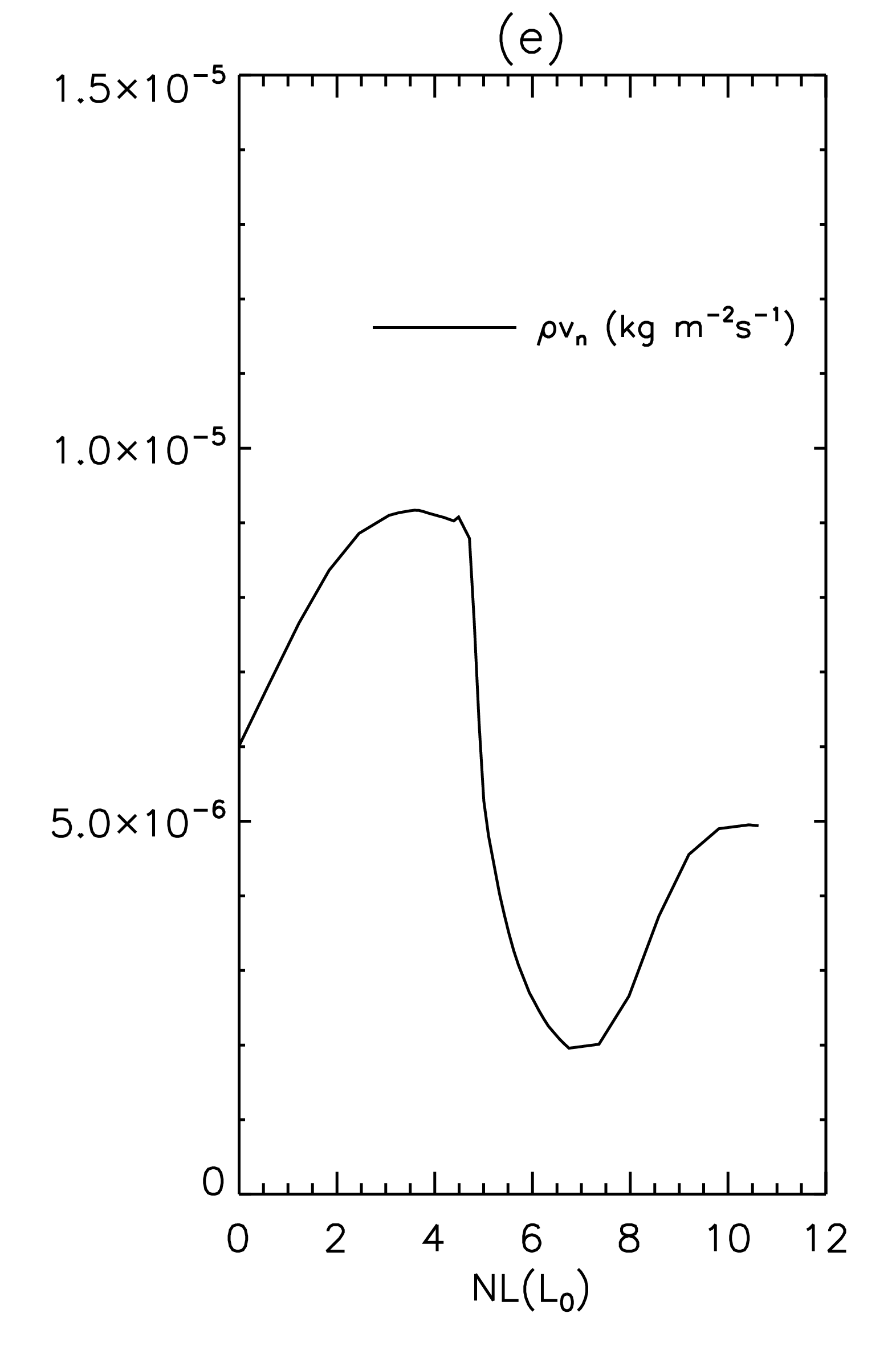}
                        \includegraphics[width=0.3\textwidth, clip=]{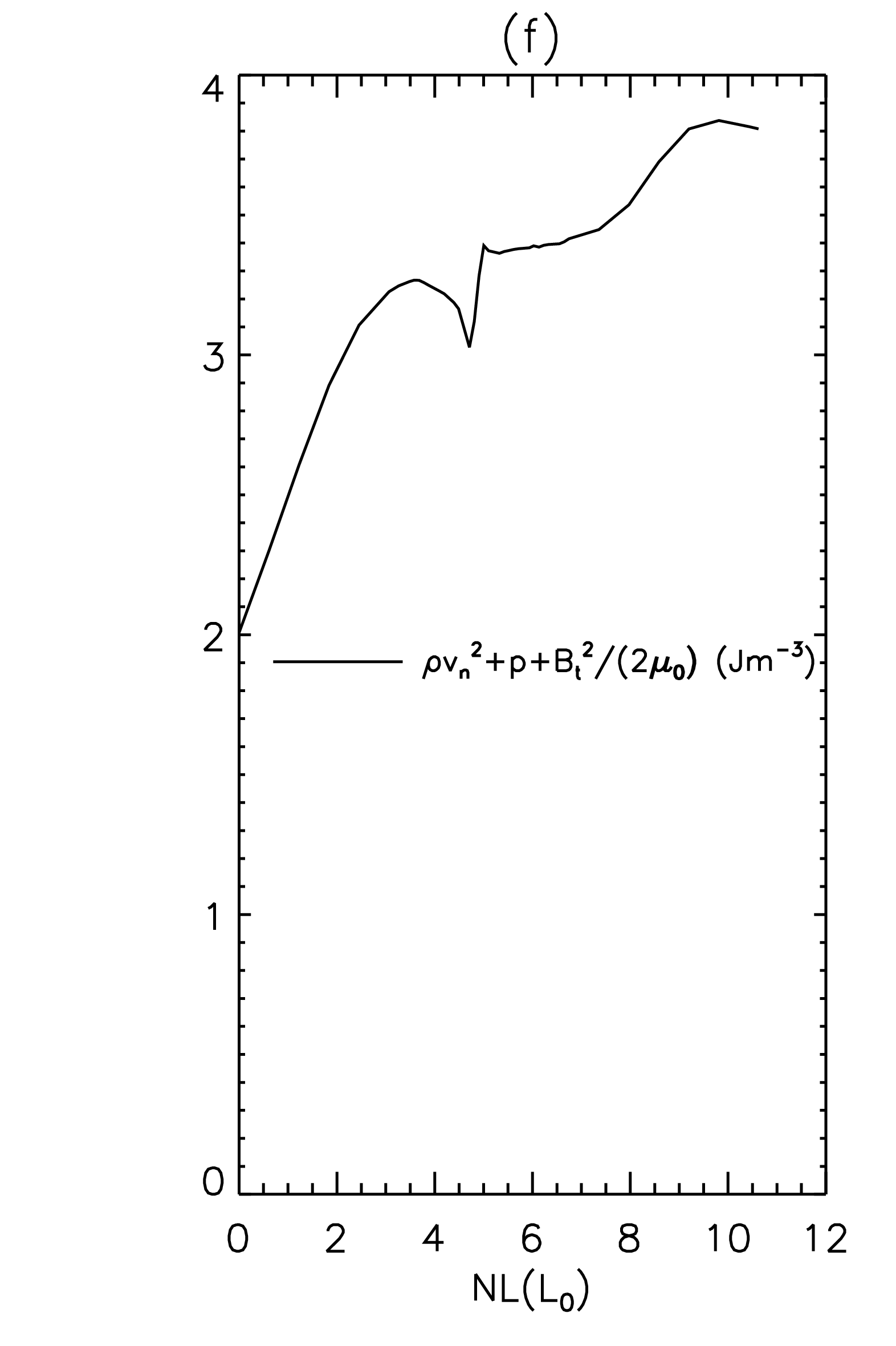}}
   \caption{(a) The color map is the divergence of the velocity field ($\nabla \cdot \mathbf{v}$) inside the box with red boundaries in the first panel of Figure~2(e), the black solid lines represent the magnetic fields and the black arrows represent the velocity vector; (b) the plasma density $\rho$  and plasma pressure p along the arrow NL in Figure~3(a); (c) the two components of the plasma velocities ($v_t$ and $v_n$) along the arrow NL; (d) the two components of the magnetic field ($B_t$ and $B_n$) along the arrow NL; (e) the variable $\rho v_n$ along the arrow NL; (f) the variable $\rho v_n^2+p+\frac{B_t^2}{2\mu_0}$ along the arrow NL. The subscript 'n' in the variables represents the component which is normal to the shock front, and the subscript 't' represents the component which is parallel to the shock front.}
 \label{fig.3}
\end{figure}

\begin{figure}
  \centerline{\includegraphics[width=0.95\textwidth, clip=]{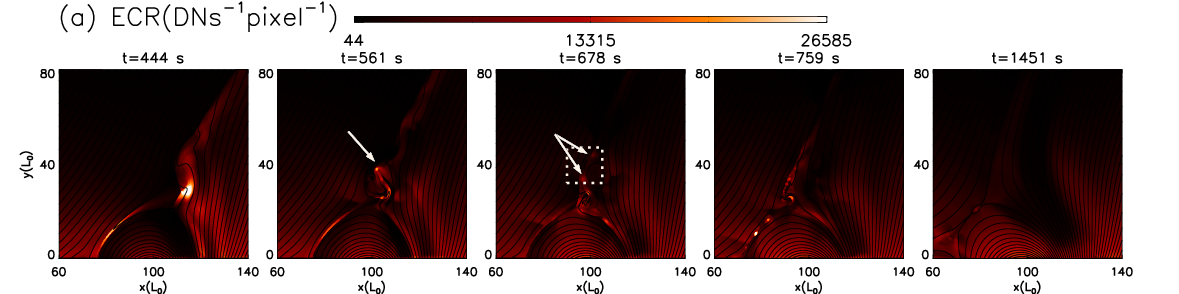}}
  \centerline{\includegraphics[width=0.29\textwidth, clip=]{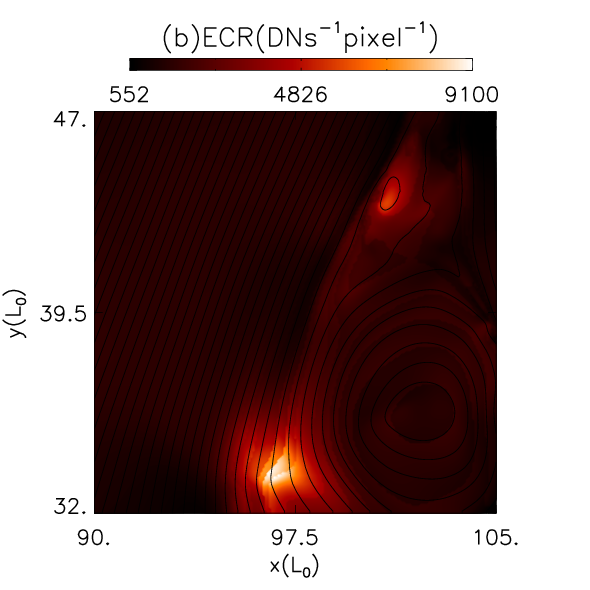}
                       \includegraphics[width=0.29\textwidth, clip=]{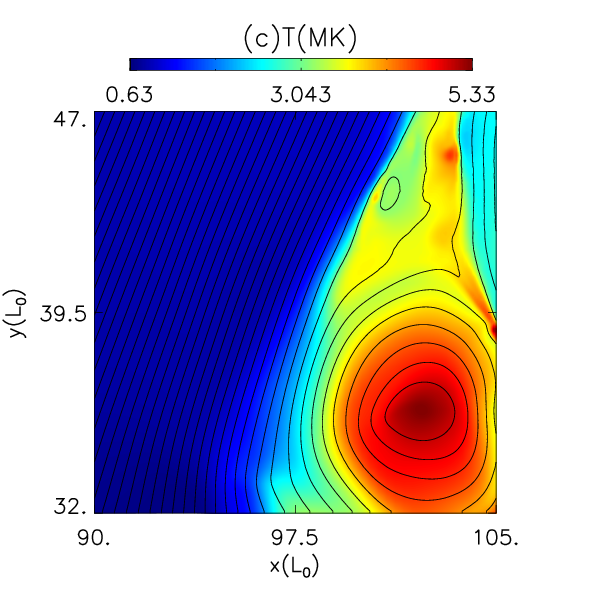}
                        \includegraphics[width=0.29\textwidth, clip=]{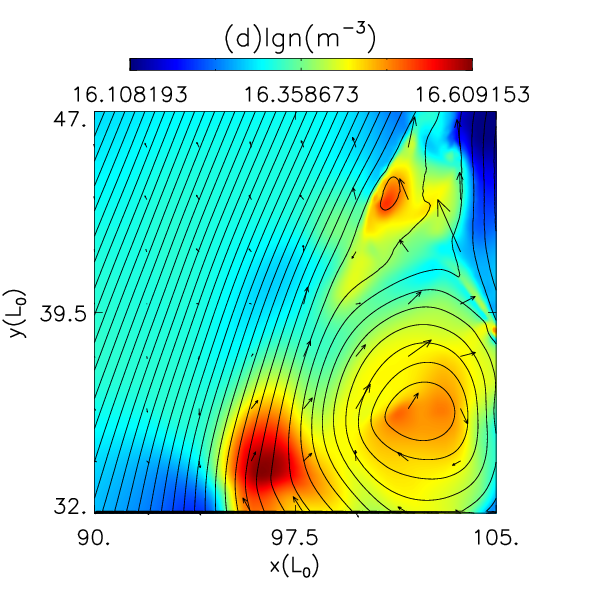}}
  \caption{(a) The distributions of the emission count rate in the AIA 335\,\AA\, channel at five different times in Case~I\@. The zoomed in region inside the white dotted box in the third panel of (a) at $t=678$ s are studied and the (b) emission count rate in the AIA 335\,\AA\, channel; (c) temperature $T$; and (d) logarithm of plasma number density $\mathrm{lg}\,n$ are shown. }
  \label{fig.4}
\end{figure}

\begin{figure}
   \centerline{\includegraphics[width=0.3\textwidth, clip=]{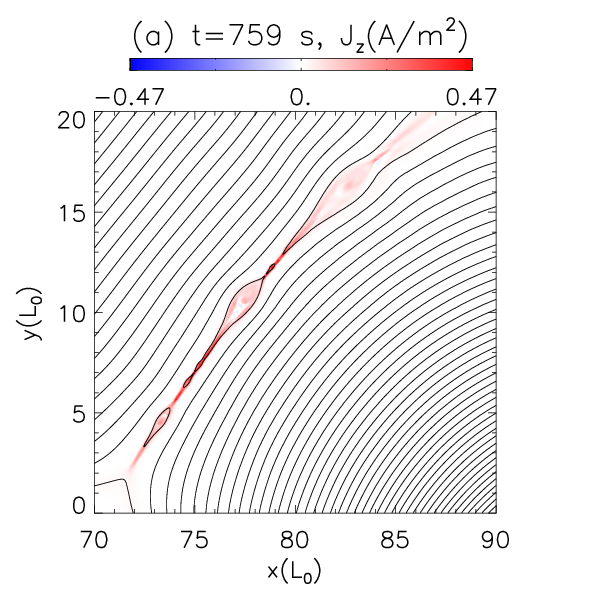}
                       \includegraphics[width=0.3\textwidth, clip=]{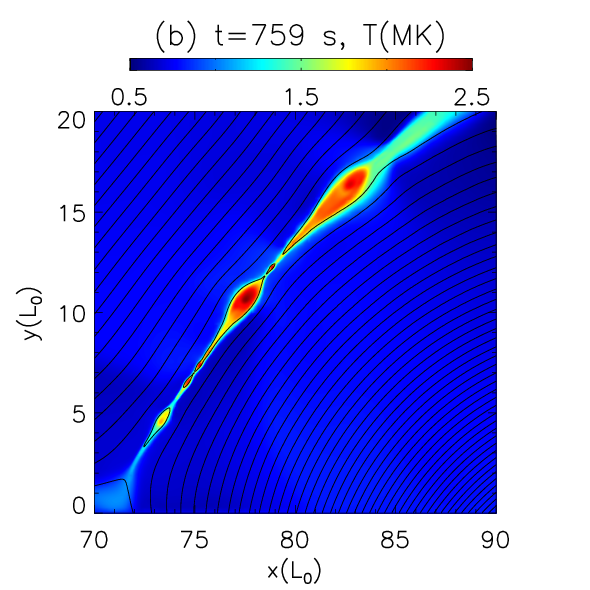}
                        \includegraphics[width=0.3\textwidth, clip=]{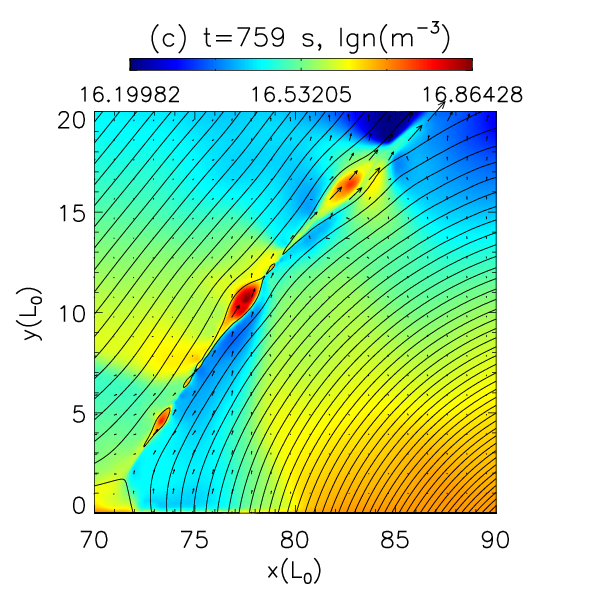}}
    \centerline{\includegraphics[width=0.3\textwidth, clip=]{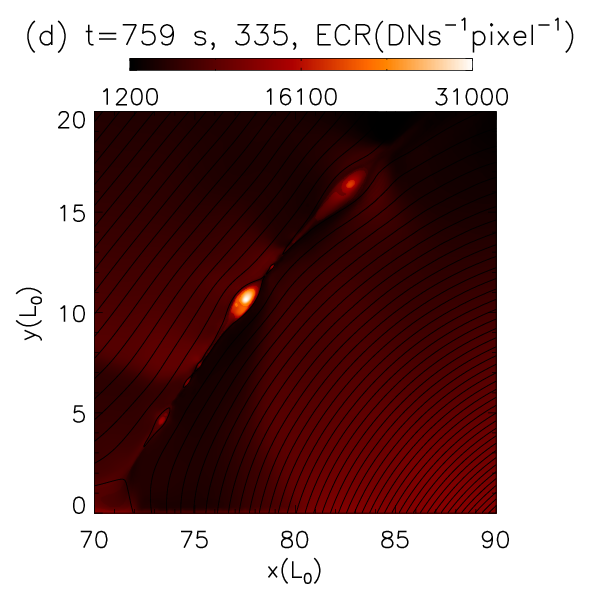}
                       \includegraphics[width=0.3\textwidth, clip=]{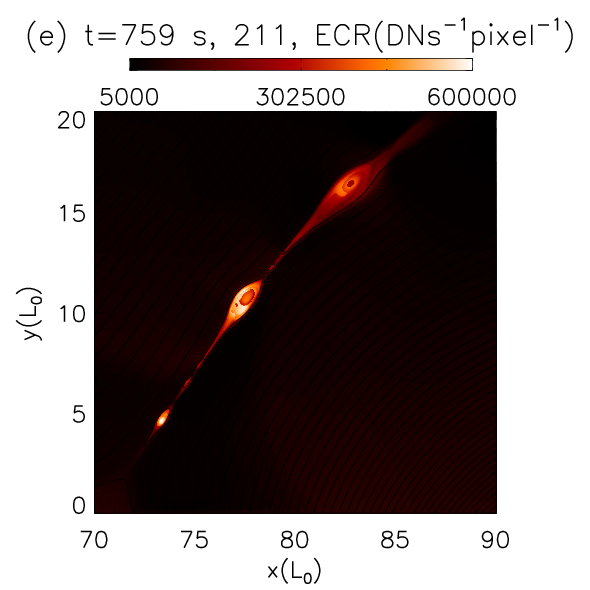}
                        \includegraphics[width=0.3\textwidth, clip=]{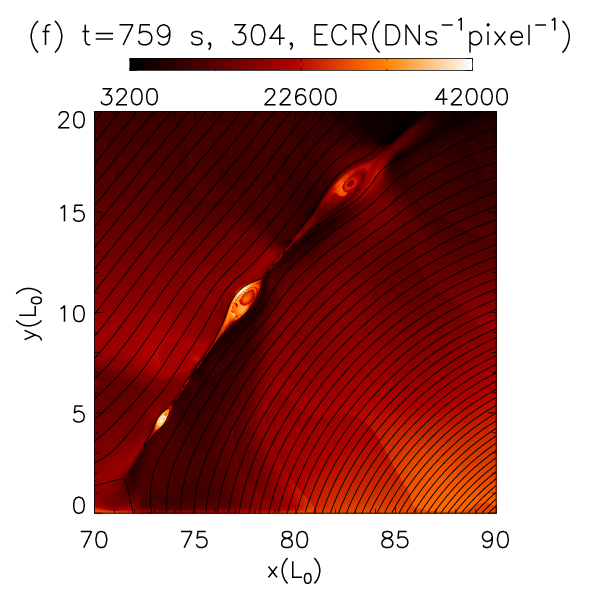}}                    
  \caption{The distributions of different variables at $t=759$~s in the main current sheet region in the bottom black dotted box in the fourth panel of Figure~2(a), (a) current density, $J_z$; (b) temperature, $T$; (c) the logarithm of plasma number density $\mathrm{lg}\,n$; (d) the emission count rate in the AIA 335\,\AA\, channel; (e) the emission count rate in the AIA 211\,\AA\, channel; (f) the emission count rate in the AIA 304\,\AA\, channel. }
  \label{fig.5}
\end{figure}

\begin{figure}
   \centerline{\includegraphics[width=0.3\textwidth, clip=]{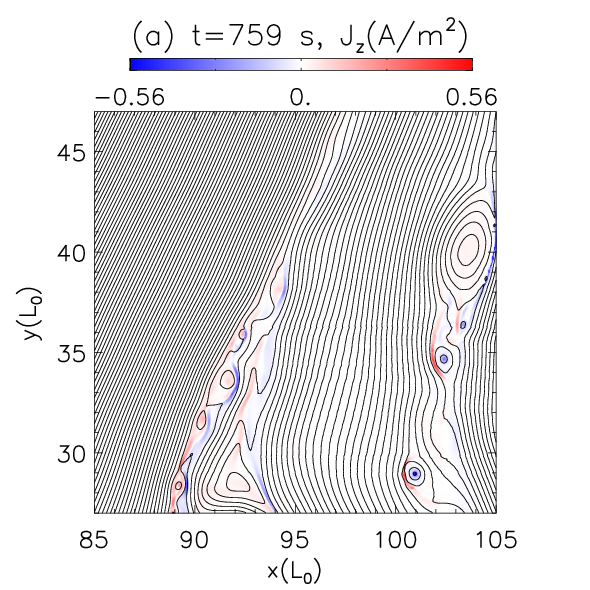}
                      \includegraphics[width=0.3\textwidth, clip=]{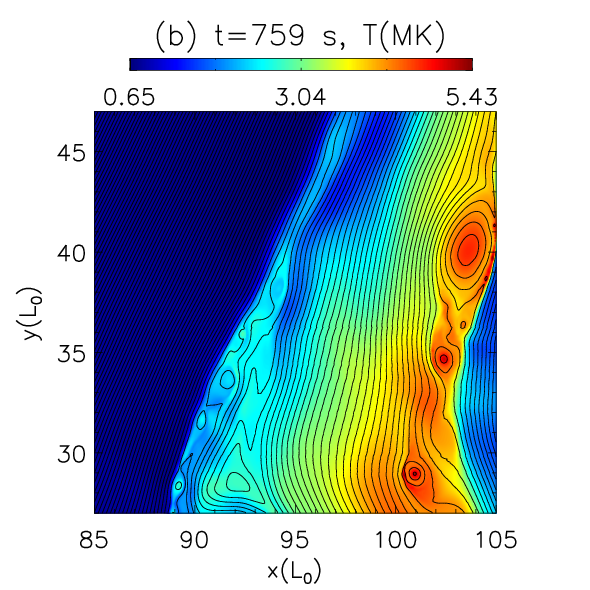}
                       \includegraphics[width=0.3\textwidth, clip=]{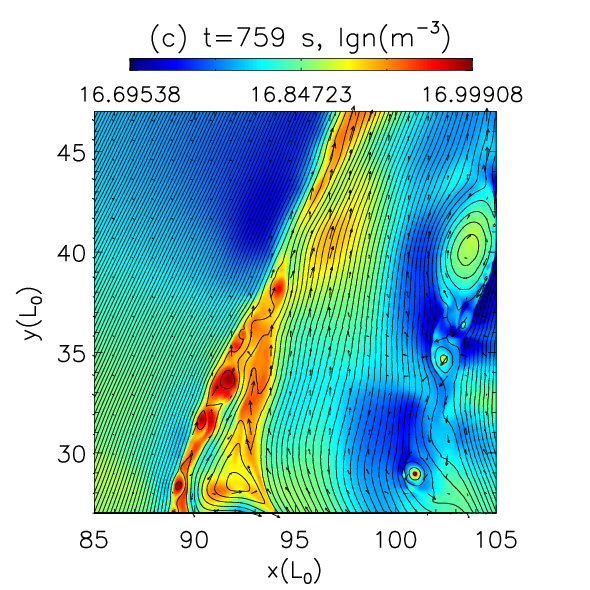}}
    \centerline{\includegraphics[width=0.3\textwidth, clip=]{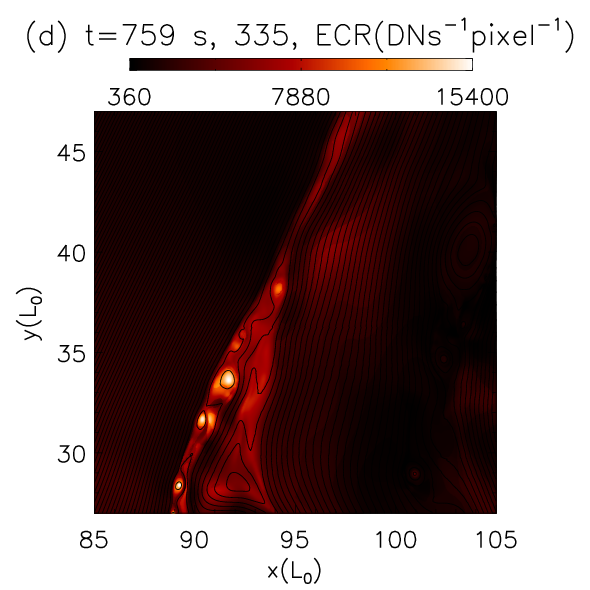}
                       \includegraphics[width=0.3\textwidth, clip=]{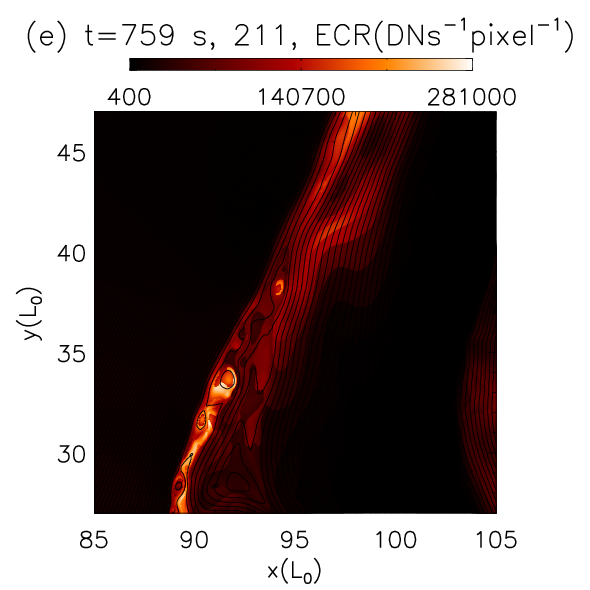}
                        \includegraphics[width=0.3\textwidth, clip=]{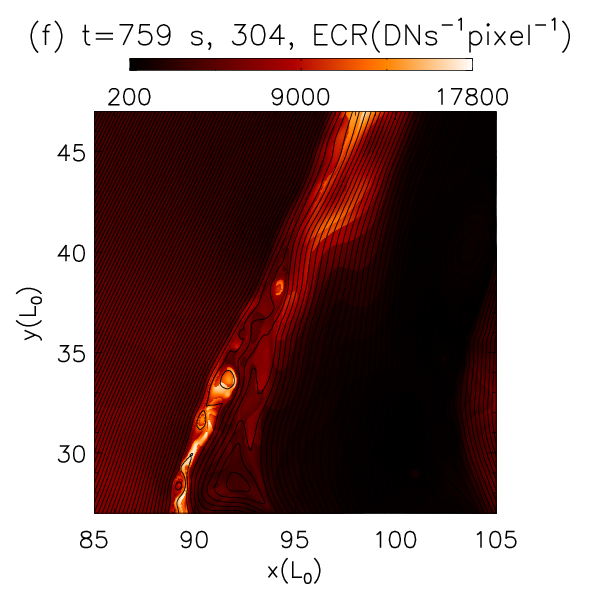}}                    
  \caption{The distributions of different variables at $t=759$~s in the jet region in the upper black dotted box in the fourth panel of Figure~ 2(a). Shown above are the (a) current density, $J_z$; (b) temperature, $T$; (c) the logarithm of plasma number density $\mathrm{lg}\,n$; (d) the emission count rate in the AIA 335\,\AA\,, channel; (e) the emission count rate in the AIA 211\,\AA\, channel; (f) the emission count rate in the AIA 304\,\AA\, channel.}
  \label{fig.6}
\end{figure}

 \begin{figure*}
     \centerline{\includegraphics[width=0.95\textwidth, clip=]{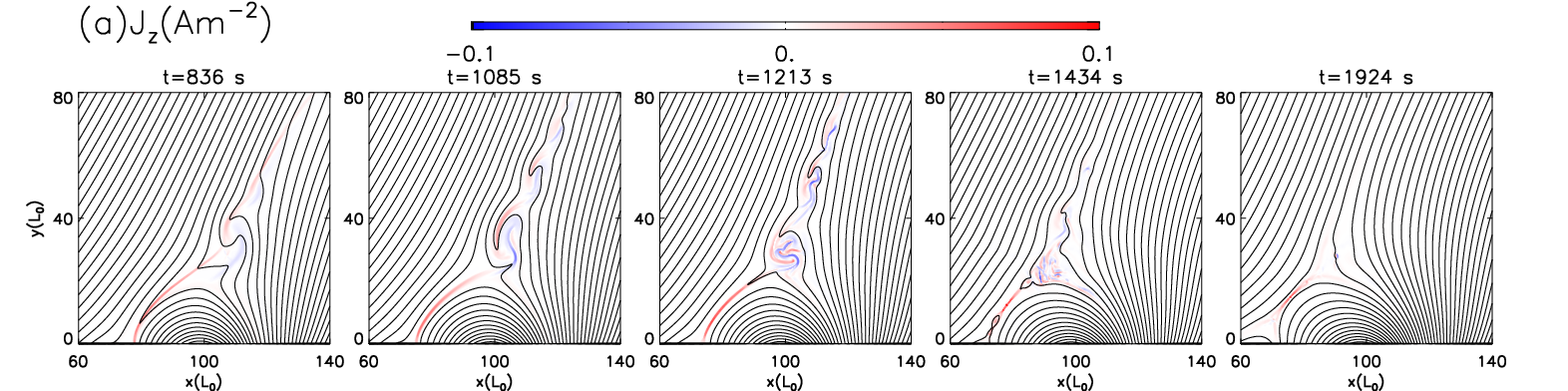}}
     \centerline{\includegraphics[width=0.95\textwidth, clip=]{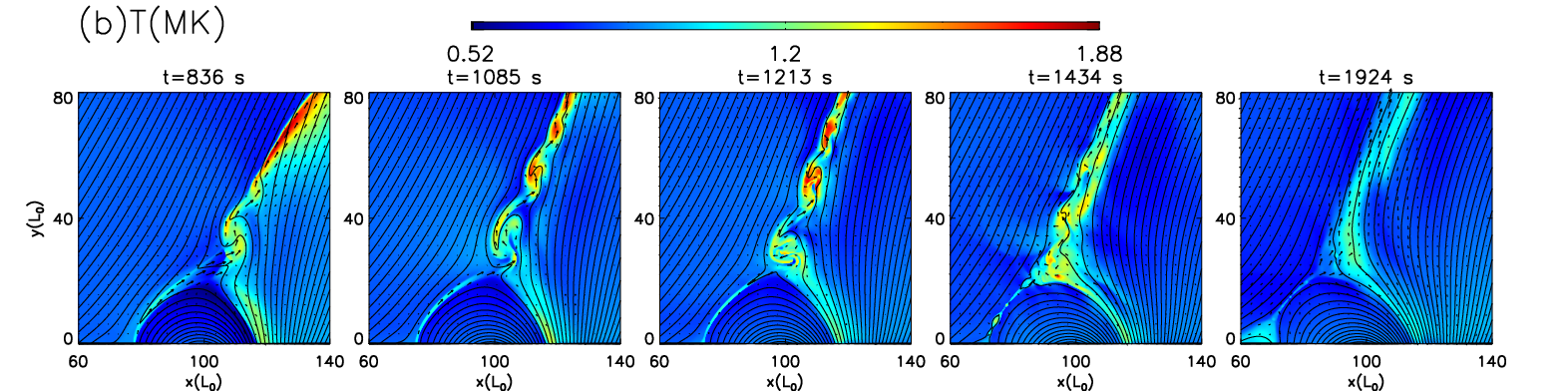}}
     \centerline{\includegraphics[width=0.95\textwidth, clip=]{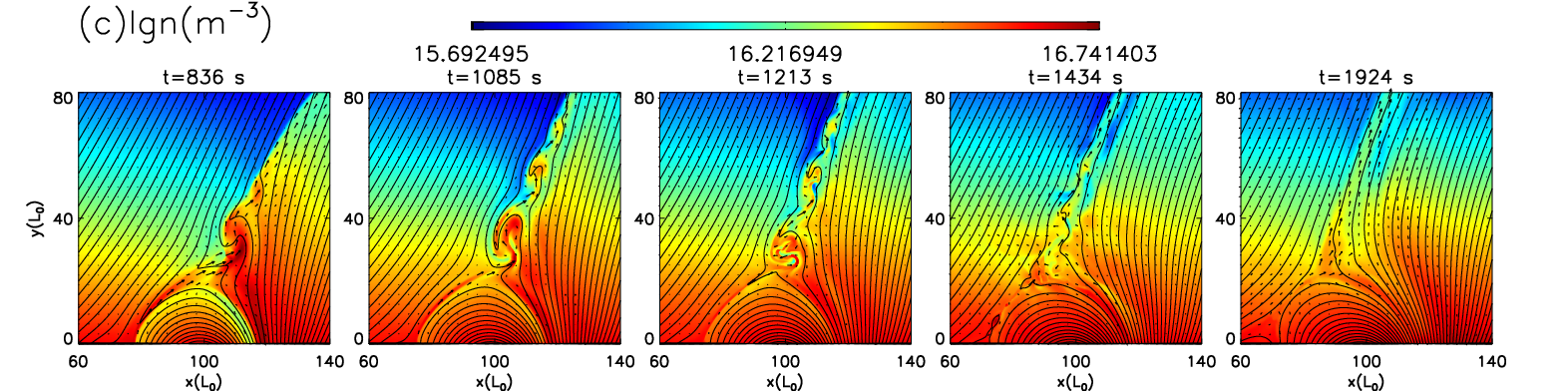}}
     \centerline{\includegraphics[width=0.95\textwidth, clip=]{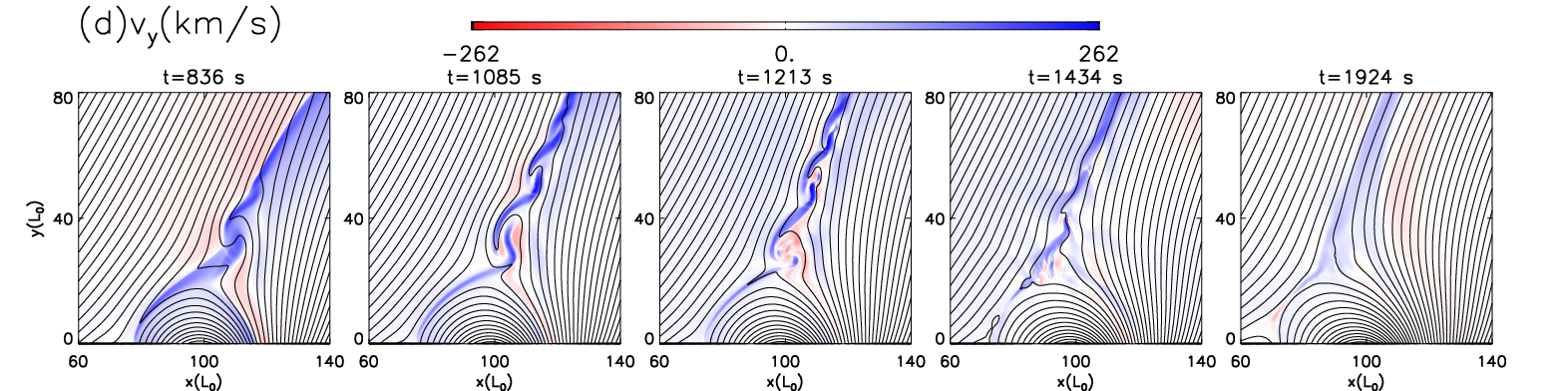}}
     \centerline{\includegraphics[width=0.95\textwidth, clip=]{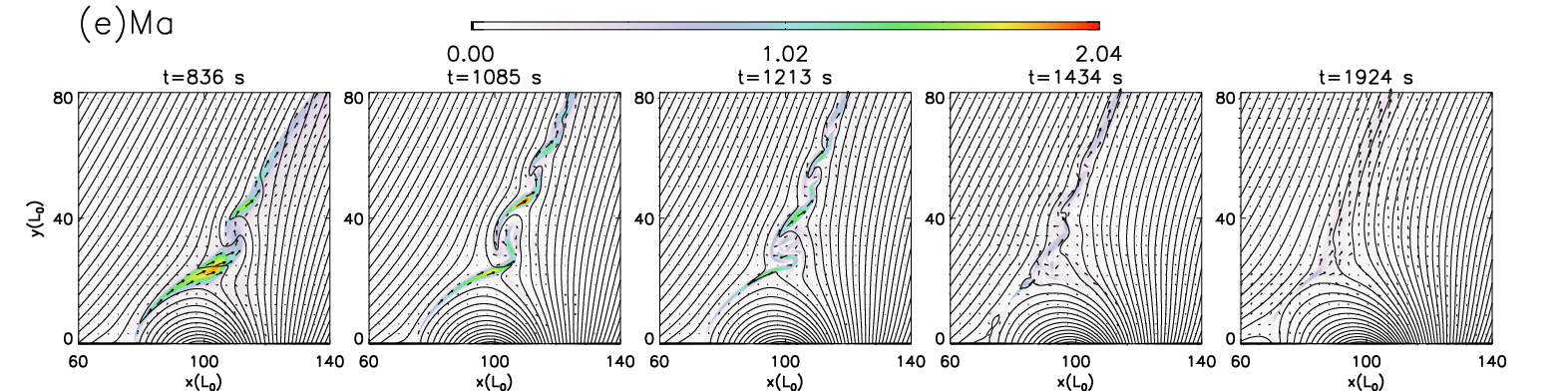}}
     \caption{The distributions of different variables at five different times in Case~II, including the (a) current density, $J_z$; (b)temperature, $T$; (c) the logarithm of plasma number density $\mathrm{lg}\,n$; (d) velocity in $y$-direction, $v_y$; (e) Mach number $Ma$. The black solid lines represent the magnetic fields and the black arrows represent the velocity in each panel. }
      \label{fig.7}
\end{figure*}

\begin{figure*}
   \centerline{\includegraphics[width=0.95\textwidth, clip=]{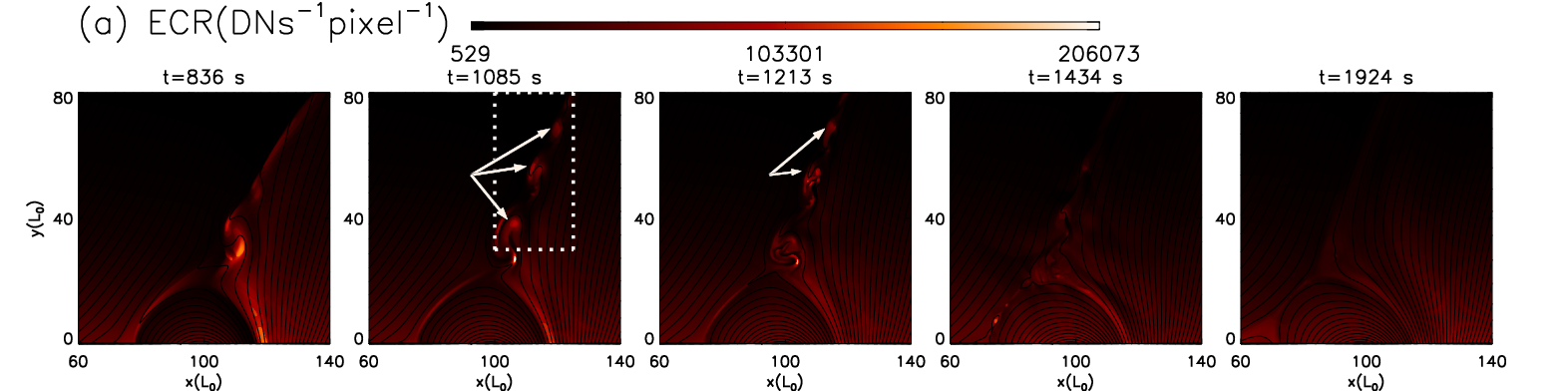}}
   \centerline{\includegraphics[width=0.17\textwidth, clip=]{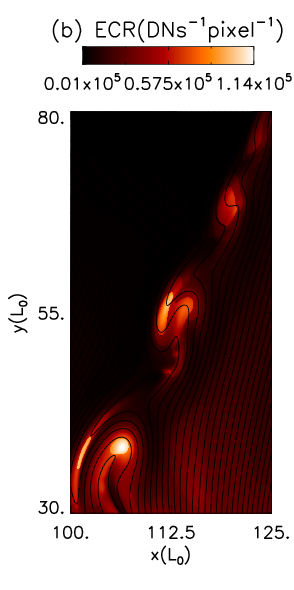}
                       \includegraphics[width=0.17\textwidth, clip=]{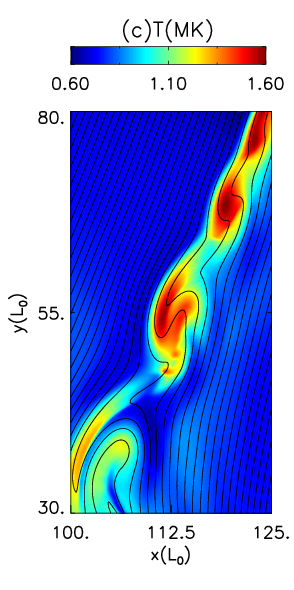}
                       \includegraphics[width=0.17\textwidth, clip=]{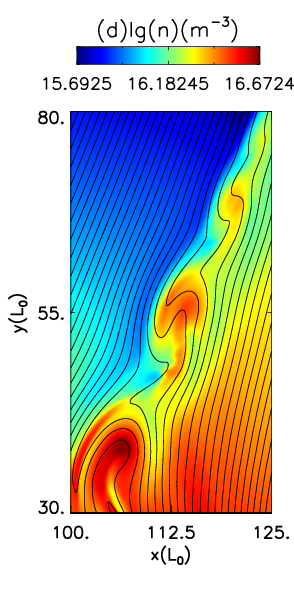}
                        \includegraphics[width=0.17\textwidth, clip=]{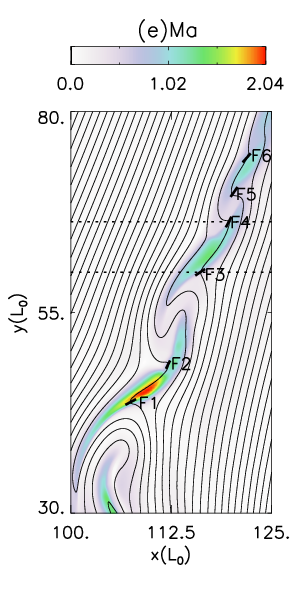}
                       \includegraphics[width=0.17\textwidth, clip=]{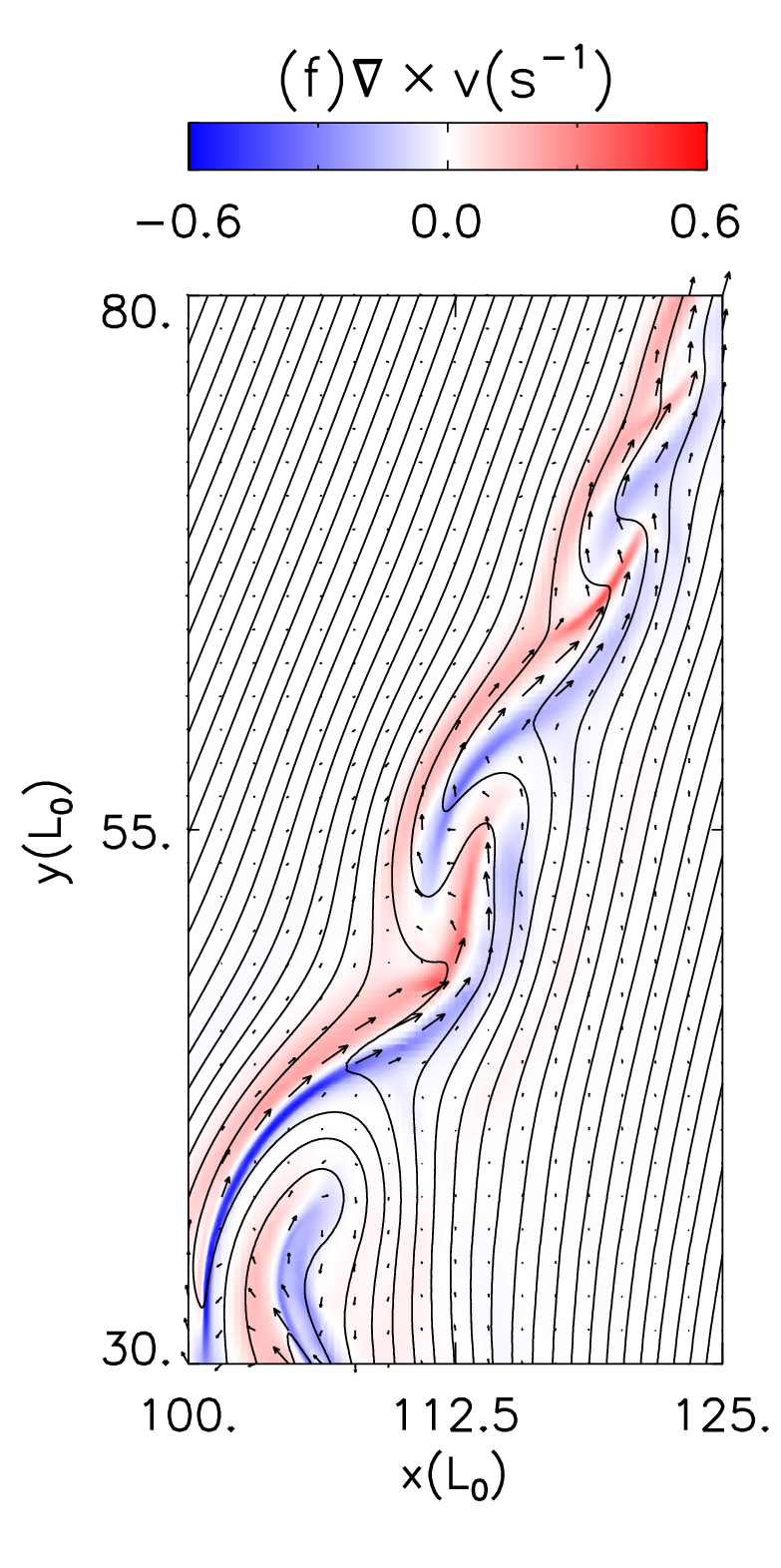}}
   \caption{(a) The distributions of the emission count rate in the AIA 211\,\AA\, channel at five different times in Case~II\@.The zoomed in region inside the white dotted box in the second panel of (a) at $t=1085$~s are studied and the (b) emission count rate in the AIA 211\,\AA\, channel; (c) temperature, $T$; (d) logarithm of plasma number density $\mathrm{lg}\,n$; (e) Mach number, $Ma$; (f) the vorticity of the velocity field are shown.}
  \label{fig.8}
\end{figure*}

\begin{figure}
  \centerline{\includegraphics[width=0.3\textwidth, clip=]{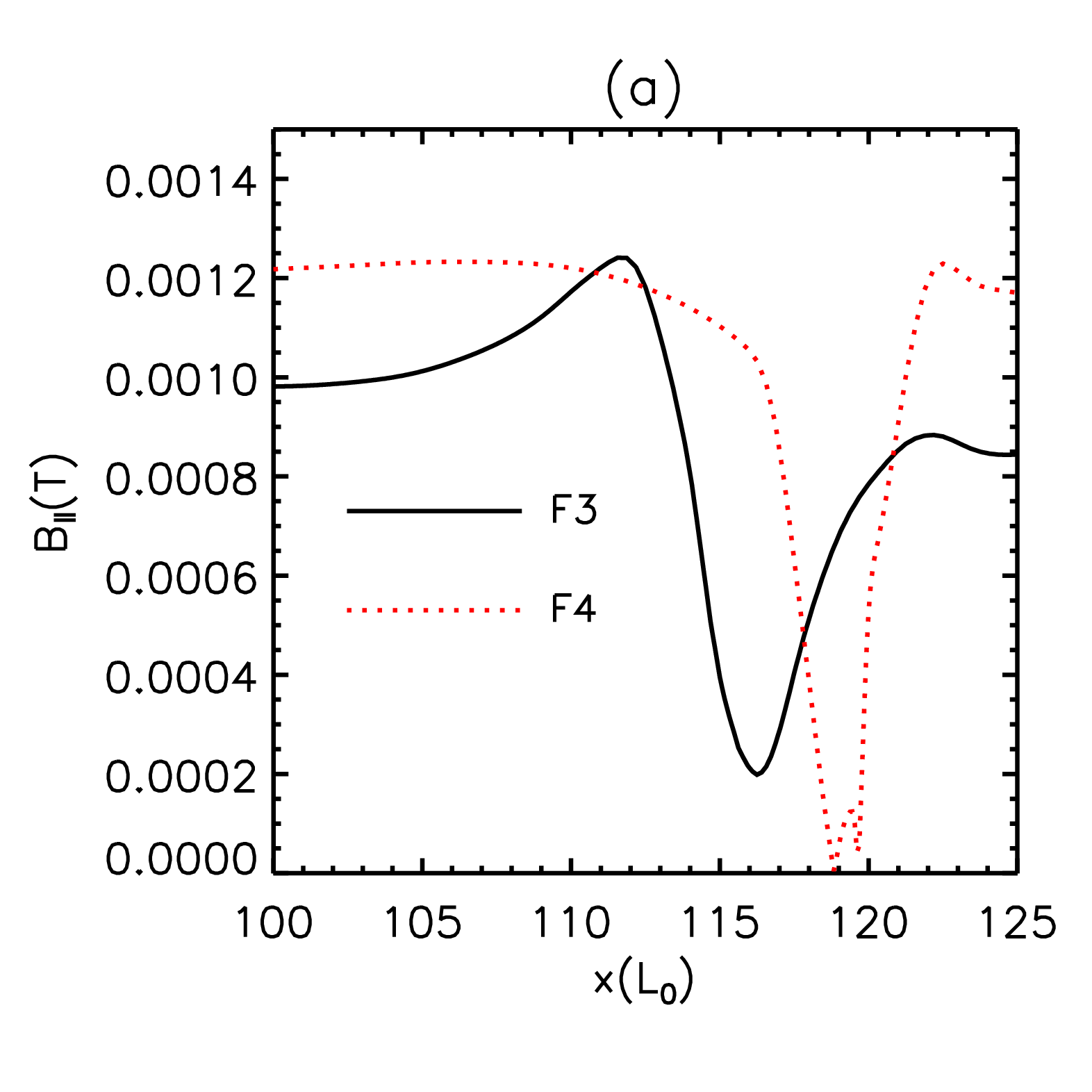}
                       \includegraphics[width=0.3\textwidth, clip=]{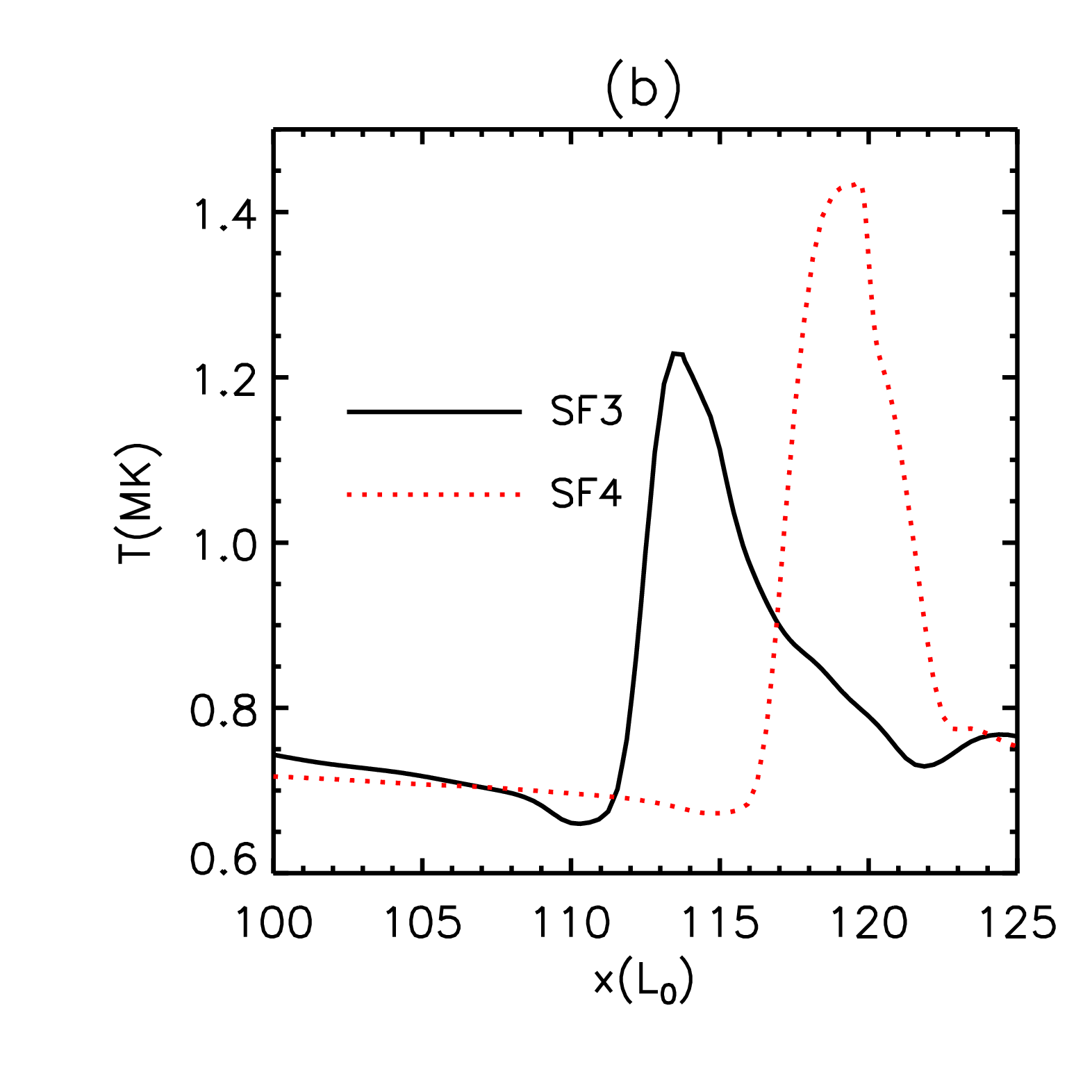}
                        \includegraphics[width=0.3\textwidth, clip=]{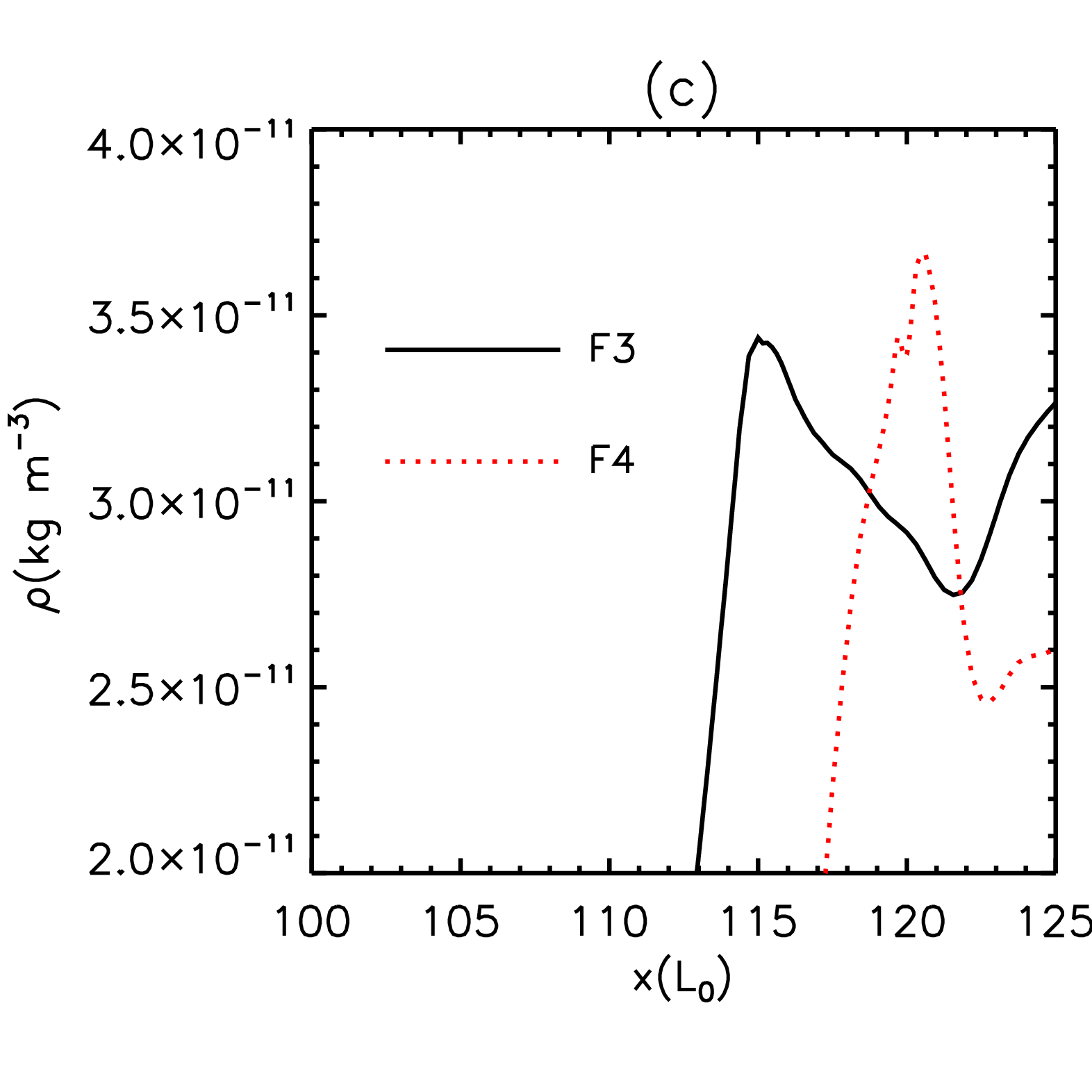}}
  \caption{The distributions of three variables along the two black dashed lines at $y=60L_0$ and $y=66.3L_0$ in Figure~8(e) are presented, the black solid lines represent the variables along $y=60L_0$ though F3 and the red dotted line represents the variables along $y=66.3L_0$ though F4. (a) The magnetic fields which are parallel to F3 and F4; (b) the temperatures; (c) the plasma densities.}
 \label{fig.9}
\end{figure}

 \begin{figure*}
    \centerline{\includegraphics[width=0.95\textwidth, clip=]{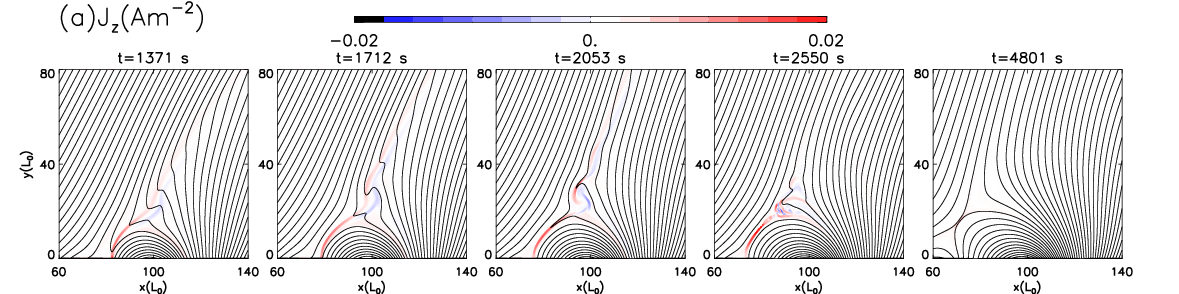}}
    \centerline{\includegraphics[width=0.95\textwidth, clip=]{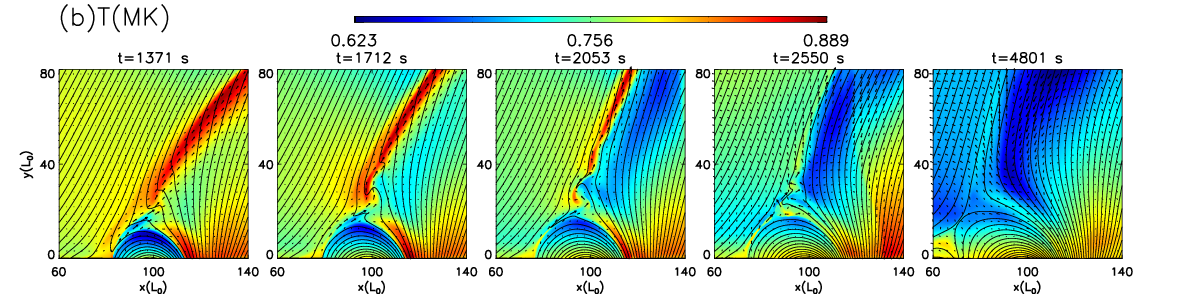}}
    \centerline{\includegraphics[width=0.95\textwidth, clip=]{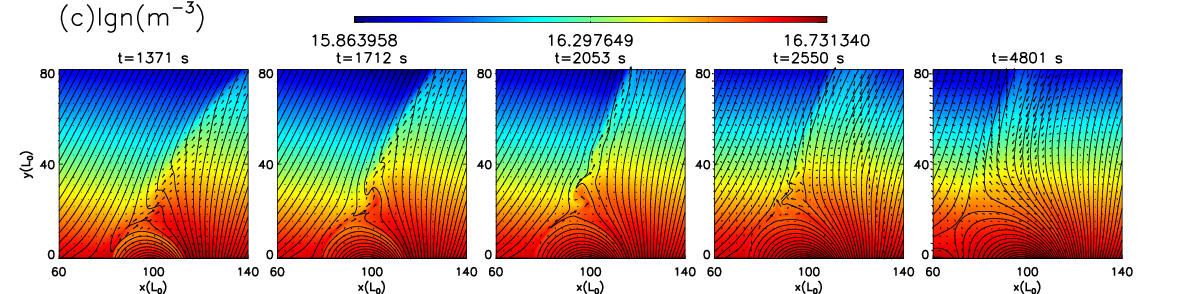}}
     \centerline{\includegraphics[width=0.95\textwidth, clip=]{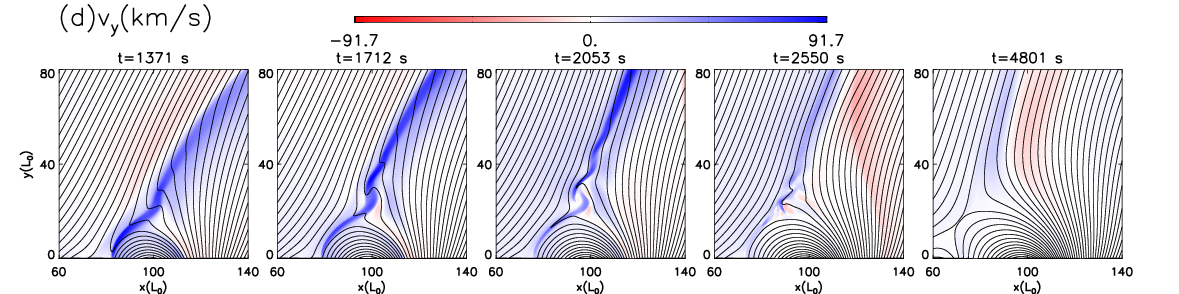}}
    \centerline{\includegraphics[width=0.95\textwidth, clip=]{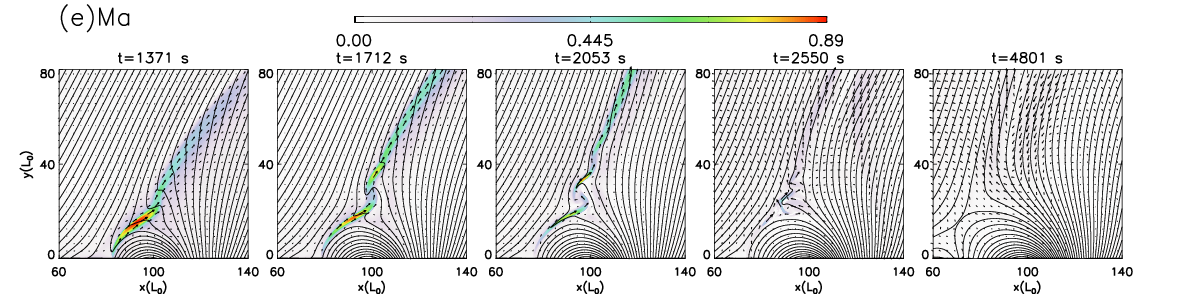}}
    \caption{The distributions of different variables at five different times in Case~III, including the (a) current density, $J_z$; (b)temperature, $T$; (c) logarithm of plasma number density $\mathrm{lg}\,n$; (d) velocity in the $y$-direction $v_y$; and (e) Mach number $Ma$. The black solid lines represent the magnetic fields and the black arrows represent the velocity in each panel. }
 \label{fig.10}
\end{figure*}

---------------------------


\begin{thebibliography}{99}
 \bibitem[Archontis et al.(2006)]{2006ApJ...645L.161A} Archontis, V., Galsgaard, K., Moreno-Insertis, F., \& Hood, A.~W.\ 2006, \apjl, 645, L161 
 \bibitem[Archontis \& Hood(2013)]{2013ApJ...769L..21A} Archontis, V., \& Hood, A.~W.\ 2013, \apjl, 769, L21
 \bibitem[Bhattacharjee et al.(2009)]{2009PhPl...16k2102B} Bhattacharjee, A., Huang, Y.-M., Yang, H., \& Rogers, B.\ 2009, Physics of Plasmas, 16, 112102
 \bibitem[Biskamp(2000)]{2000mrp..book.....B} Biskamp, D.\ 2000, Magnetic reconnection in plasmas, Cambridge, UK: Cambridge University Press, 2000 xiv, 387 p.~Cambridge monographs on plasma physics, vol.~3, ISBN 0521582881.
 \bibitem[Chae et al.(1999)]{1999ApJ...513L..75C} Chae, J., Qiu, J., Wang, H., \& Goode, P.~R.\ 1999, \apjl, 513, L75 
 \bibitem[Chen \& Shibata(2000)]{2000ApJ...545..524C} Chen, P.~F., \& Shibata, K.\ 2000, \apj, 545, 524 
 \bibitem[Chen et al.(2015)]{2015Sci...350.1238C} Chen, B., Bastian, T.~S., Shen, C., et al.\ 2015, Science, 350, 1238 
 \bibitem[Chen et al.(2015)]{2015ApJ...815...71C} Chen, J., Su, J., Yin, Z., et al.\ 2015, \apj, 815, 71 
 \bibitem[Del Zanna et al.(2015)]{2015A&A...582A..56D} Del Zanna, G., Dere, K.~P., Young, P.~R., Landi, E., \& Mason, H.~E.\ 2015, \aap, 582, A56 
 \bibitem[Chifor et al.(2008)]{2008A&A...491..279C} Chifor, C., Isobe, H., Mason, H.~E., et al.\ 2008, \aap, 491, 279
 \bibitem[Comisso \& Bhattacharjee(2016)]{2016JPlPh..82f5901C} Comisso, L., \& Bhattacharjee, A.\ 2016, Journal of Plasma Physics, 82, 595820601
 \bibitem[De Pontieu et al.(2014)]{2014SoPh..289.2733D} De Pontieu, B., Title, A.~M., Lemen, J.~R., et al.\ 2014, \solphys, 289, 2733 
 \bibitem[Dere et al.(2009)]{2009A&A...498..915D} Dere, K.~P., Landi, E., Young, P.~R., et al.\ 2009, \aap, 498, 915
  \bibitem[Ding et al.(2010)]{2010A&A...510A.111D} Ding, J.~Y., Madjarska, M.~S., Doyle, J.~G., \& Lu, Q.~M.\ 2010, \aap, 510, A111 
 \bibitem[Fan(2001)]{2001ApJ...554L.111F} Fan, Y.\ 2001, \apjl, 554, L111 
 \bibitem[Forbes \& Priest(1984)]{1984SoPh...94..315F} Forbes, T.~G., \& Priest, E.~R.\ 1984, \solphys, 94, 315 
 \bibitem[Guo et al.(2014)]{2014ApJ...796L..29G} Guo, L.-J., Huang, Y.-M., Bhattacharjee, A., \& Innes, D.~E.\ 2014, \apjl, 796, L29
 \bibitem[Keppens et al.(1999)]{1999JPlPh..61....1K} Keppens, R., T{\'o}th, G., Westermann, R.~H.~J., \& Goedbloed, J.~P.\ 1999, Journal of Plasma Physics, 61, 1
 \bibitem[Klimchuk et al.(2008)]{2008ApJ...682.1351K} Klimchuk, J.~A., Patsourakos, S., \& Cargill, P.~J.\ 2008, \apj, 682, 1351-1362
 \bibitem[Jiang et al.(2012)]{2012ApJ...751..152J} Jiang, R.-L., Fang, C., \& Chen, P.-F.\ 2012, \apj, 751, 152
 \bibitem[Lee et al.(2013)]{2013ApJ...766....1L} Lee, K.-S., Innes, D.~E., Moon, Y.-J., et al.\ 2013, \apj, 766, 1 
 \bibitem[Lemen et al.(2012)]{2012SoPh..275...17L} Lemen, J.~R., Title, A.~M., Akin, D.~J., et al.\ 2012, \solphys, 275, 17 
 \bibitem[Loureiro et al.(2007)]{2007PhPl...14j0703L} Loureiro, N.~F., Schekochihin, A.~A., \& Cowley, S.~C.\ 2007, Physics of Plasmas, 14, 100703
 \bibitem[Magara(2001)]{2001ApJ...549..608M} Magara, T.\ 2001, \apj, 549, 608 
 \bibitem[Mart{\'{\i}}nez-Sykora et al.(2008)]{2008ApJ...679..871M} Mart{\'{\i}}nez-Sykora, J., Hansteen, V., \& Carlsson, M.\ 2008, \apj, 679, 871-888 
 \bibitem[Moreno-Insertis et al.(2008)]{2008ApJ...673L.211M} Moreno-Insertis, F., Galsgaard, K., \& Ugarte-Urra, I.\ 2008, \apjl, 673, L211 
 \bibitem[Moreno-Insertis \& Galsgaard(2013)]{2013ApJ...771...20M} Moreno-Insertis, F., \& Galsgaard, K.\ 2013, \apj, 771, 20 
 \bibitem[Mulay et al.(2016)]{2016A&A...589A..79M} Mulay, S.~M., Tripathi, D., Del Zanna, G., \& Mason, H.\ 2016, \aap, 589, A79
 \bibitem[Nagai(1980)]{1980SoPh...68..351N} Nagai, F.\ 1980, \solphys, 68, 351
 \bibitem[Ni et al.(2010)]{2010PhPl...17e2109N} Ni, L., Germaschewski, K., Huang, Y.-M., et al.\ 2010, Physics of Plasmas, 17, 052109 
 \bibitem[Ni et al.(2013)]{2013PhPl...20f1206N} Ni, L., Lin, J., \& Murphy, N.~A.\ 2013, Physics of Plasmas, 20, 061206
 \bibitem[Ni et al.(2015)]{2015ApJ...799...79N} Ni, L., Kliem, B., Lin, J., \& Wu, N.\ 2015, \apj, 799, 79 
 \bibitem[Ni et al.(2015)]{2015ApJ...812...92N} Ni, L., Lin, J., Mei, Z., \& Li, Y.\ 2015, \apj, 812, 92 
 \bibitem[Ni et al.(2016)]{2016ApJ...832..195N} Ni, L., Lin, J., Roussev, I.~I., \& Schmieder, B.\ 2016, \apj, 832, 195
 \bibitem[Nishizuka et al.(2008)]{2008ApJ...683L..83N} Nishizuka, N., Shimizu, M., Nakamura, T., et al.\ 2008, \apjl, 683, L83 
 \bibitem[Nistic{\`o} et al.(2009)]{2009SoPh..259...87N} Nistic{\`o}, G., Bothmer, V., Patsourakos, S., \& Zimbardo, G.\ 2009, \solphys, 259, 87 
 \bibitem[N{\'o}brega-Siverio et al.(2016)]{2016ApJ...822...18N} N{\'o}brega-Siverio, D., Moreno-Insertis, F., \& Mart{\'{\i}}nez-Sykora, J.\ 2016, \apj, 822, 18
 \bibitem[Pariat et al.(2009)]{2009ApJ...691...61P} Pariat, E., Antiochos, S.~K., \& DeVore, C.~R.\ 2009, \apj, 691, 61
 \bibitem[Pariat et al.(2015)]{2015A&A...573A.130P} Pariat, E., Dalmasse, K., DeVore, C.~R., Antiochos, S.~K., \& Karpen, J.~T.\ 2015, \aap, 573, A130
 \bibitem[Priest(2014)]{2014masu.book.....P} Priest, E.\ 2014, Magnetohydrodynamics of the Sun, by Eric Priest, Cambridge, UK: Cambridge University Press, 2014, p177-188
 \bibitem[Savcheva et al.(2007)]{2007PASJ...59S.771S} Savcheva, A., Cirtain, J., Deluca, E.~E., et al.\ 2007, \pasj, 59, S771
 \bibitem[Schmieder et al.(2013)]{2013A&A...559A...1S} Schmieder, B., Guo, Y., Moreno-Insertis, F., et al.\ 2013, \aap, 559, A1  
 \bibitem[Shen et al.(2011)]{2011ApJ...735L..43S} Shen, Y., Liu, Y., Su, J., \& Ibrahim, A.\ 2011, \apjl, 735, L43
 \bibitem[Shen et al.(2012)]{2012ApJ...745..164S} Shen, Y., Liu, Y., Su, J., \& Deng, Y.\ 2012, \apj, 745, 164
 \bibitem[Shibata et al.(1992)]{1992PASJ...44L.173S} Shibata, K., Ishido, Y., Acton, L.~W., et al.\ 1992, \pasj, 44, L173 
 \bibitem[Shibata et al.(2007)]{2007Sci...318.1591S} Shibata, K., Nakamura, T., Matsumoto, T., et al.\ 2007, Science, 318, 1591 
 \bibitem[Shimojo et al.(1996)]{1996PASJ...48..123S} Shimojo, M., Hashimoto, S., Shibata, K., et al.\ 1996, \pasj, 48, 123
 \bibitem[Tian \& Chen(2016)]{2016ApJ...824...60T} Tian, C., \& Chen, Y.\ 2016, \apj, 824, 60 
 \bibitem[Tian et al.(2014)]{2014Sci...346A.315T} Tian, H., DeLuca, E.~E., Cranmer, S.~R., et al.\ 2014, Science, 346, 1255711 
 \bibitem[Wyper et al.(2016)]{2016ApJ...827....4W} Wyper, P.~F., DeVore, C.~R., Karpen, J.~T., \& Lynch, B.~J.\ 2016, \apj, 827, 4 
  \bibitem[Yang et al.(2013)]{2013ApJ...777...16Y} Yang, L., He, J., Peter, H., et al.\ 2013, \apj, 777, 16
  \bibitem[Yokoyama \& Shibata(1995)]{1995Natur.375...42Y} Yokoyama, T., \& Shibata, K.\ 1995, \nat, 375, 42
  \bibitem[Yokoyama \& Shibata(1996)]{1996PASJ...48..353Y} Yokoyama, T., \& Shibata, K.\ 1996, \pasj, 48, 353 
  \bibitem[Zenitani \& Miyoshi(2011)]{2011PhPl...18b2105Z} Zenitani, S., \& Miyoshi, T.\ 2011, Physics of Plasmas, 18, 022105 
  \bibitem[Zhang \& Ji(2014)]{2014A&A...567A..11Z} Zhang, Q.~M., \& Ji, H.~S.\ 2014, \aap, 567, A11 
  \bibitem[Zhang et al.(2016)]{2016SoPh..291..859Z} Zhang, Q.~M., Ji, H.~S., \& Su, Y.~N.\ 2016, \solphys, 291, 859 
  \bibitem[Zhang \& Zhang(2017)]{2017ApJ...834...79Z} Zhang, Y., \& Zhang, J.\ 2017, \apj, 834, 79
   \bibitem[Ziegler(2008)]{2008CoPhC.179..227Z} Ziegler, U.\ 2008, Computer Physics Communications, 179, 227
   \bibitem[Ziegler(2011)]{2011JCoPh.230.1035Z} Ziegler, U.\ 2011, Journal of Computational Physics, 230, 1035
 \end{thebibliography}
\end{document}